\documentclass{aa}  
\usepackage{graphicx}
\usepackage{txfonts}
\usepackage{natbib}
\bibpunct{(}{)}{;}{a}{}{,} 

\def\aap{A\&A}
\def\apj{ApJ}
\def\apjs{ApJS}
\def\aj{AJ}

\def \hi {\ion{H}{i}~}

\def\kms{km\,s$^{-1}$}
\def\kmss{km\,s$^{-1}~$}
\def\msun{M$_{\sun}$}

\def\rsun{R$_{\sun}$}
\def\rsuns{R$_{\sun}~$}

\def\vsuns{v$_{\sun}~$}
\def\deg{\hbox{$^\circ$}}

\begin{document}
   \title{Global properties of the \hi distribution in the outer Milky Way.}

   \subtitle{Planar and extra-planar gas }

   \author{P.\,M.\,W. Kalberla \& L. Dedes}

   \institute{ Argelander-Institut f\"ur Astronomie, Universit\"at Bonn,
     Auf dem H\"ugel 71, 53121 Bonn,
     Germany\\
     \email{pkalberla@astro.uni-bonn.de,ldedes@astro.uni-bonn.de} }
      
   \authorrunning{P.\,M.\,W. Kalberla \& L. Dedes\ } 

   \titlerunning{\hi distribution in the outer Milky Way}

   \offprints{P.\,M.\,W. Kalberla}

   \date{Received December 13 2007 / Accepted March 31 2008 }

  \abstract 
  {The determination of the global structure of the planar and extra-planar
    Milky Way \hi disk depends critically on a reliable database but also on
    reasonable assumptions about the shape of the Milky Way rotation curve.}
  {We derive the 3-D \hi volume density distribution for the Galactic disk out
    to $R \sim 60$ kpc. }
  {Our analysis is based on parameters for the warp and rotation curve derived
    previously. The data are taken from the
    Leiden/Argentine/Bonn all sky 21-cm line survey. }
  {The Milky Way \hi disk is significantly warped but shows a coherent
    structure out to $R \sim 35$ kpc. The radial surface density distribution,
    the densities in the middle of the warped plane, and the \hi scale
    heights all follow exponential relations. The radial scale length for the
    surface density distribution of the \hi disk is 3.75 kpc. Gas at the
    outskirts for $ 40 \la R \la 60$ kpc is described best by a distribution
    with an exponential radial scale length of 7.5 kpc and a velocity
    dispersion of 74 \kms. Such a highly turbulent medium fits also well with 
    the average shape of the high velocity profile wings observed at high
    latitudes. The turbulent pressure gradient of such extra-planar gas is
    on average in balance with the gravitational forces. About 10\% of the
    Milky Way \hi gas is in this state. The large scale \hi distribution is
    lopsided; for $R \ga 15$ kpc there is more gas in the south. The \hi
    flaring indicates that this asymmetry is caused by a dark matter wake,
    located at $R \sim 25$ kpc in direction of the Magellanic System. }
  {The \hi disk is made up of two major components. Most prominent is the
    normal \hi disk which can be traced to $R \sim 35$ kpc. This is surrounded
    by a patchy distribution of highly turbulent gas reaching large scale
    heights but also large radial distances. At the position of the Sun the
    exponential scale height in the $z$ direction is 3.9 kpc.  This component
    resembles the anomalous gas discovered previously in some galaxies. }
  \maketitle
%

\section{Introduction}

The gaseous content of galaxies is a crucial counterpart to the stellar
population but has the additional advantage that it can be traced far beyond
the stellar population, radially but also perpendicular to the disk. Stars are
born out of gas condensations and their deaths affect the three-dimensional
evolution of the Galactic disc. The gas traces such activities. Here we are
interested in the steady state situation caused by dynamical processes as
described e.g. by \citet{deAvillez2000}. The 21-cm line traces these
processes, in particular the diffuse part of the interstellar medium (ISM). A
thin disk is expected in the plane, overlaid by a thick \hi disk which is
associated with an ionized layer. Depending on the supernova rate in the
underlying disc a more or less established disc-halo interface is expected.

\begin{figure*}[!ht]
   \centering
\includegraphics[width=18cm]{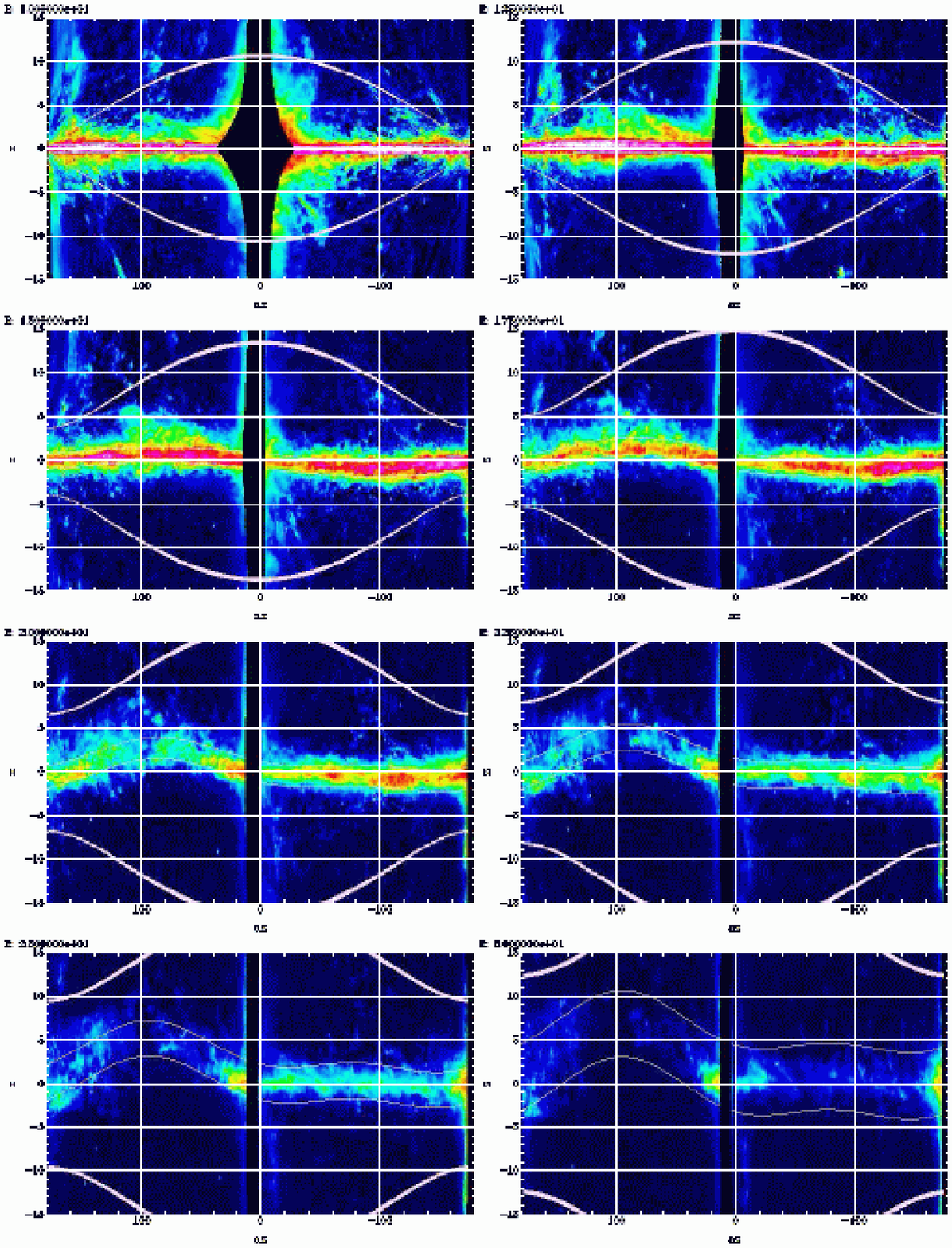} 
   \caption{Derived volume density distribution at $R =$ 10, 12.5, 15, 17.5,
     20, 22.5, 25, 30 kpc (top to bottom).  Locations affected by local
     emission with $|v_{lsr}| < 10$ \kmss are flagged dark. We display
     densities $ n < 0.5$ cm$^{-3}$. A logarithmic transfer function was
     chosen to emphasize low densities.  The thin white lines indicate the
     average flaring at a $3 \sigma$ level and enclose the warped disk. The
     thick white lines indicate latitudes $|b| = 30 \deg \pm 0.5 \deg$.  }
         \label{Fig_R_cuts}
\end{figure*}

Most prominent, comparing the gaseous content of galaxies with the stellar
distribution, is the large radial extent of the gaseous disk. The outer part
is only slightly affected by the stellar disk but traces environmental
influences such as influences from the intergalactic medium. We 
explore global parameters for layers above the disk and gas at the outskirts
of the Milky Way.  For such an analysis we need to determine the properties of
the dominant part of the \hi distribution first, i.e. the \hi disk which
co-rotates with the stellar disk. Knowing the properties of the gas within the
disk we may proceed to determine the nature of the gas around this disk.

In Sect. 2 we briefly introduce the database and the conversion of brightness
temperatures $T_B(l,b,v)$ to densities $n(R,z,\phi)$. The observer's location
inside the Galactic disk causes some ambiguities and constraints that need to
be discussed. In Sect. 3 we derive radial dependencies of the average surface
density distribution within the disk. This is followed in Sect. 4 by a
discussion of average properties of the mid-plane volume densities. We find
that the gas distribution within the Milky Way \hi disk is approximately
exponential for a large part of the disk. The gas properties at $R \ga 35$ kpc
differ significantly; this is studied in Sect. 5. At the outskirts we find
that the \hi distribution needs to be described as a highly turbulent phase.
In Sect. 6 we consider the \hi distribution at the polar caps. We require
hydrostatic equilibrium and derive a velocity dispersion of $\sigma = 74$ \kms
for the halo gas phase. Such a phase explains not only the \hi distribution at
the outskirts but also the average local emission high above the disk.  About
10\% of the \hi is in this phase and we demonstrate that the Milky Way would
look very similar to NGC 891 or NGC 2403 if it were observed edge-on at large
distances. Sect. 7 contains a summary.


\section{Database and derived density distribution}

Our analysis is based on the Leiden/Argentine/Bonn (LAB) \hi line survey
\citep{Kalberla2005}. This survey combines the southern sky survey of the
Instituto Argentino de Radioastronom\'ia (IAR) \citep{Bajaja2005} with an
improved version of the Leiden/Dwingeloo Survey (LDS)
\citep{Atlas1997}. Currently, this is the most sensitive Milky Way \hi line
survey with the most extensive coverage both spatially and kinematically. The
line profiles have been corrected for spurious sidelobe emission; we expect
therefore that profile wings, most critical for our investigations, 
are only mildly contaminated by instrumental problems.

\subsection{The density distribution $n(R,z,\phi)$}

All the individual steps necessary to convert the observed brightness
temperature distribution $T_B(l,b,v)$ as a function of galactic longitude,
latitude and LSR velocity to densities $n(R,z,\phi)$ have been described by
\citet{Kalberla2007}.  We use cylindrical coordinates $R, z, \phi$ to describe
the Milky Way disk; $ \phi= 0 $ is in direction $l = 0$. We use the IAU
recommendations for the galactic constants; \rsuns = 8.5 kpc and \vsuns = 220
\kms.

Figure \ref{Fig_R_cuts} shows several global properties of the Galactic \hi
density distribution. We have chosen a common logarithmic transfer function to
cover the high dynamic range of the density fluctuations within the Galactic
disk. The average extension of the flaring disk layer is indicated. The thin
white lines indicate the thickness of the disk at a $3 \sigma$ level of the
second moment fitted by \citet{Kalberla2007} . The warp is obvious for $R \ga
15$ kpc as a sine-wave pattern. With increasing distance, the southern part of
the disk, $-180\deg \la \phi \la 0\deg$, returns to $z \sim 0\deg$ while the
northern part, $0\deg \la \phi \la 180\deg$, bends up strongly and becomes
increasingly disrupted. Within the galactic plane several shells are
detectable; some of them have very low central densities. All kind of spurs,
arcs, filaments, and chimneys are visible, indicating violent interactions for
most parts of the disk, most pronounced for $R \la 20$ kpc.
 
We do not discuss details visible in Fig. \ref{Fig_R_cuts} but a few comments
concerning the $T_B(l,b,v)$ to $n(R,z,\phi)$ conversion are necessary. The
shape of the Milky Way rotation curve, but also the location of the Sun within
the Galactic plane, causes complex geometrical transformations. A white line
is plotted in Fig. \ref{Fig_R_cuts} at latitudes $|b| = 30\deg \pm 0.5\deg$
for visualization. Some ambiguities are discussed next.

\begin{figure}[!ht]
   \centering
   \includegraphics[width=9cm]{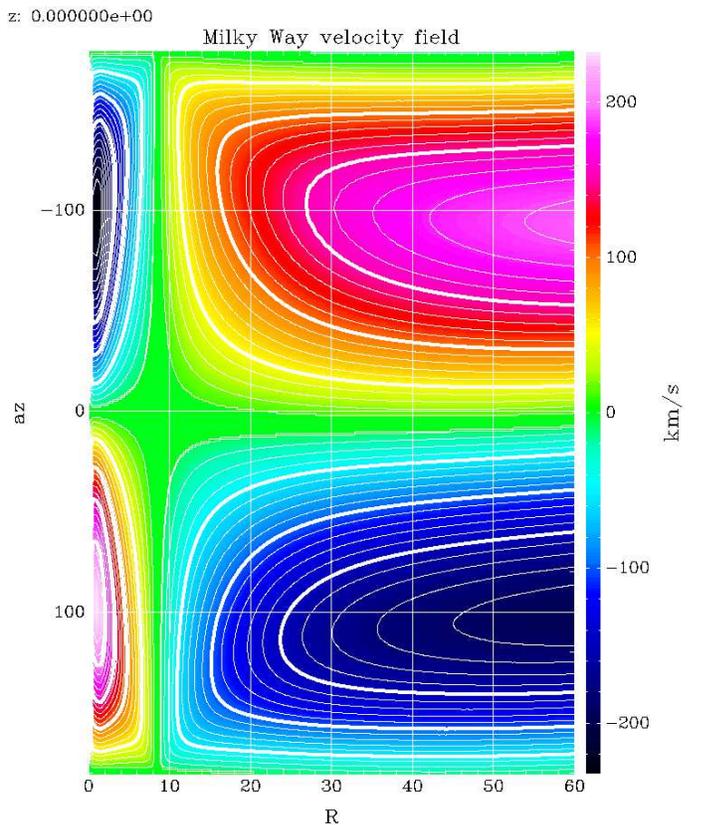}
   \caption{Milky Way velocity field at mid-plane for the mass distribution
     derived by \citet{Kalberla2007}. The isophotes indicate velocities $ -190
     < v_{lsr} < 190$ \kms in steps of 10 \kms. The thick lines are at $|
     v_{lsr}| = 50$, 100 and 150 \kms.   }
    \label{Fig_vel}
\end{figure}

Local gas at low velocities may cause spurious features in the density
distribution, most serious for $|v_{\rm LSR}| < 10 $ \kms. These regions have
been disregarded, indicated by the black color in
Fig. \ref{Fig_R_cuts}. Figure \ref{Fig_vel} displays the velocity field at
mid-plane. The central green region, enclosed by isophotes, corresponds to
$|v_{\rm LSR}| < 10 $ \kms. There are some more regions with unreliable
kinematic distances. In the direction of the Galactic center, we discarded
data originating from $|R| < 3.5 $ kpc. This region, approximately the extent
of the bar, is affected by ambiguities due to large velocity dispersions of
the emission lines \citep[][Fig. 8]{Weiner99}. Low $dv/dR$ values,
predominantly at $ |\phi| \la 30\deg$, $\phi \ga 150\deg$, and $\phi \la
-150\deg$ introduce additional distance uncertainties. We strictly exclude
these regions from further analysis and use for the northern part of the
Galactic plane the range $ 30\deg < \phi < 150\deg$ and $ 210\deg < \phi <
330\deg$ for the southern part.

A fraction of the observed \hi gas distribution is presumably not associated
with the disk. Intermediate and high velocity clouds (IVCs and HVCs,
respectively) are most probably located above the disk \citep{HVCbook}.  One
might think that such features should be excluded from the database prior to
the determination of $n(R,z,\phi)$ for the \hi disk. We emphasize that no such
filtering was applied to the observations.

The scale height of the \hi disk and the mid-plane density was determined by
adaptive filtering of the volume density distribution. The basic underlying
assumption is that the disk can be characterized as a homogeneous gas layer in
circular rotations, for details we refer to \citet{Kalberla2007}. The thin
white lines in Fig.  \ref{Fig_R_cuts} characterize the extension of the gas in
the Galactic plane.  We display the $3 \sigma$ level of the second moment
fitted by \citet{Kalberla2007} to the flaring gas layer. The $3 \sigma$ limit
is somewhat arbitrary, a different measure might be chosen to distinguish
between disk and extra-planar gas. Whatever criteria might be applied, Fig.
\ref{Fig_R_cuts} shows that most of the \hi gas settles down in a thin disk
phase while there remains an extra-planar contribution, either diffuse or
filamentary, but in most cases unambiguously associated with the disk. This
extra-planar part was excluded when calculating disk scale heights and the
mid-plane densities, but was left in the database for the rest of the
analysis. There is direct evidence that on average 5\% of the \hi gas is
located outside the disk (Fig. \ref{Fig_fract}). As discussed later, this is a
lower limit only.

\begin{figure}[!ht]
   \centering
   \includegraphics[angle=-90,width=9cm]{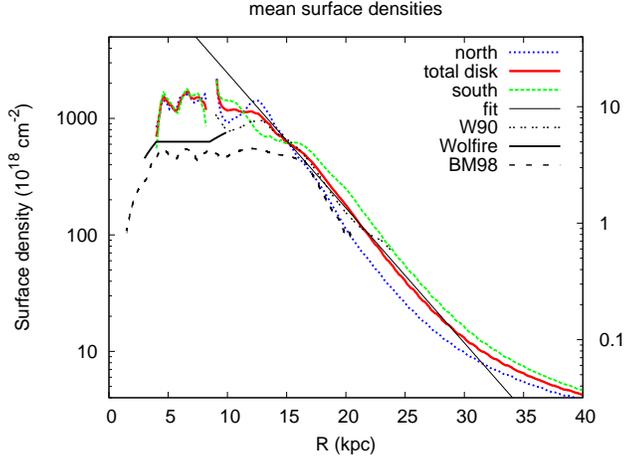}
   \caption{Derived mean surface densities of the \hi gas perpendicular to the
     disk (red). The dotted blue line gives $N_H(R)$ in the northern, the
     dotted green line in the southern part of the Milky Way. The dashed black
     line reproduces the distribution published by \citet[][BM98]{BM} in
     Fig. 9.19. \citet{Wolfire2003} (thick black line) used partial data from
     \citet[][W90]{Wouterloot90}. }
         \label{Fig_surf}
   \end{figure}

\begin{figure}[!h]
   \centering
   \includegraphics[angle=-90,width=9cm]{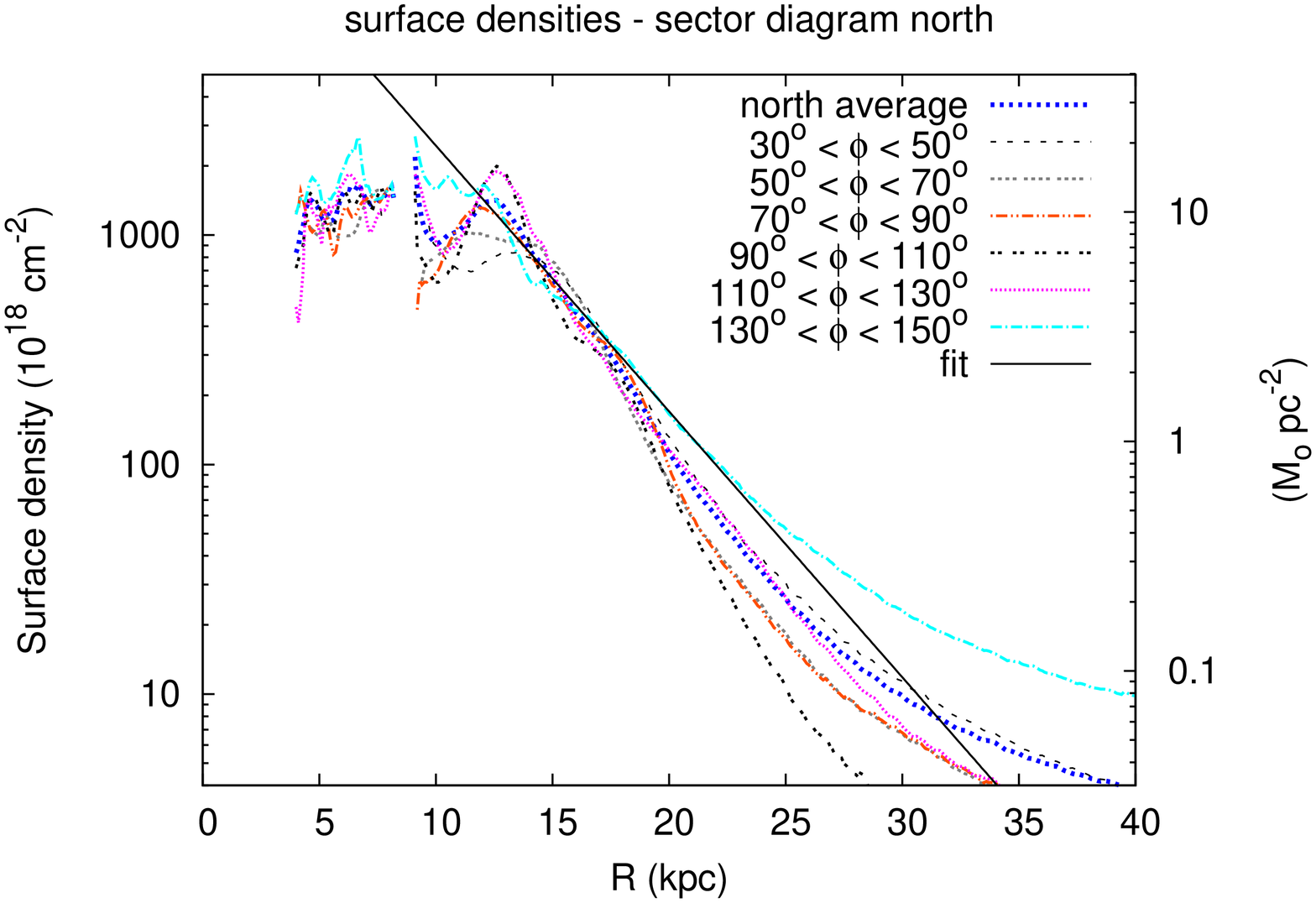}
   \includegraphics[angle=-90,width=9cm]{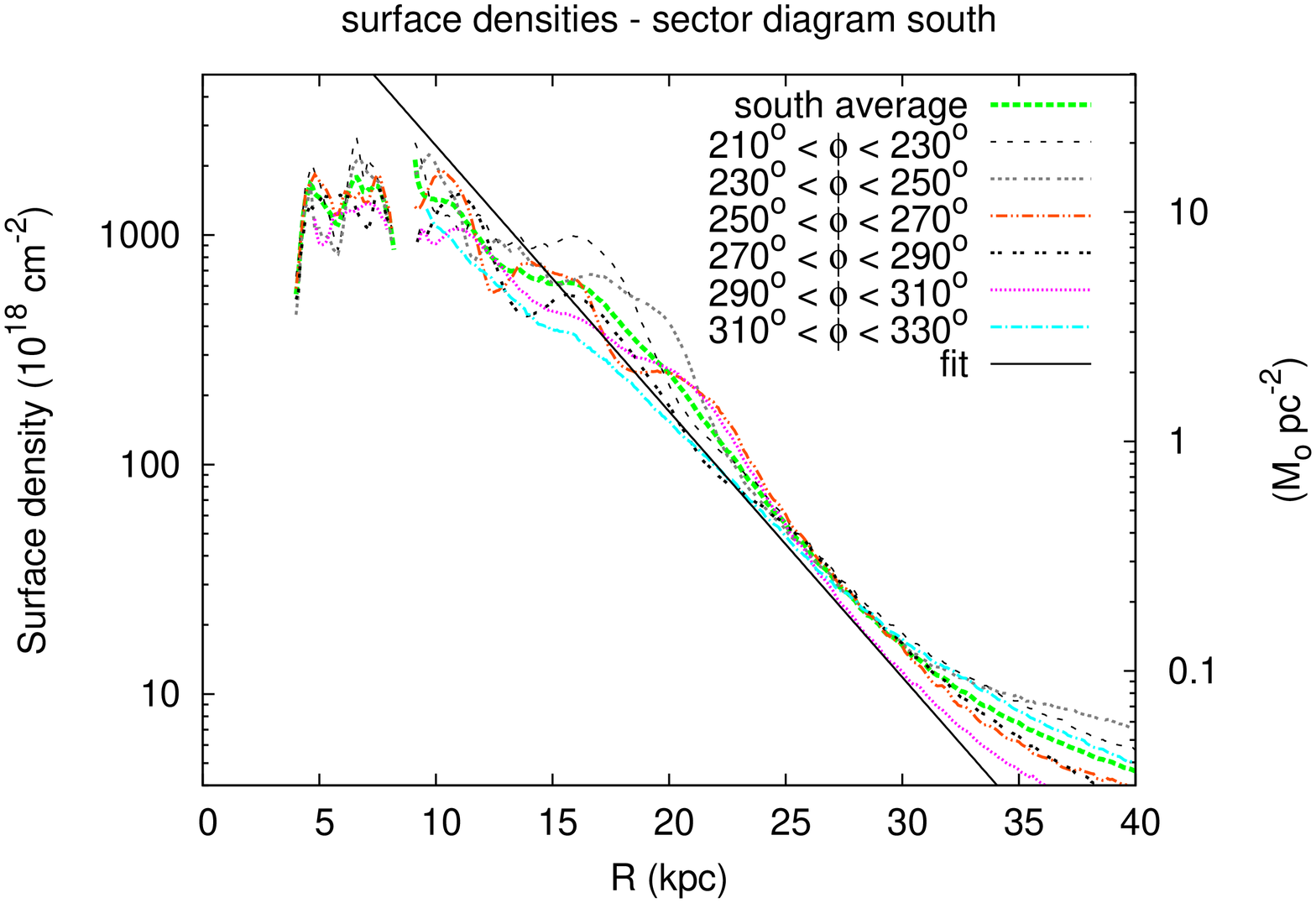}
   \caption{Systematic variations of the average \hi surface density
     distribution determined in $20\deg$ wide sectors for $30\deg < \phi <
     150\deg$ (north) and $210\deg < \phi < 330\deg$ (south). For comparison
     with Fig. \ref{Fig_surf} the fit (black) and total averages for north
     and south (blue and green respectively) are given. 
}
         \label{Fig_sdens_scatter}
\end{figure}

\begin{figure}[!h]
   \centering
   \includegraphics[angle=-90,width=9cm]{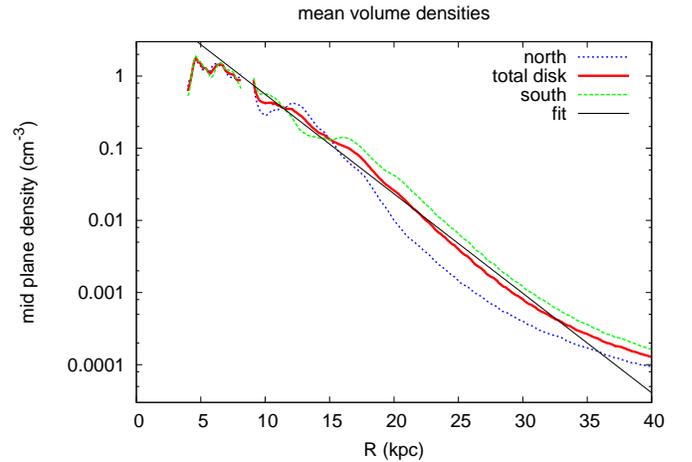}
   \caption{Derived average mid-plane volume density of the \hi gas (red). The
     dotted blue line gives $n(R)$ in the north ($30\deg < \phi < 150\deg$) ,
     the dotted green line in the southern part of the Milky Way ($210\deg <
     \phi < 330\deg$). The black line represents an exponential fit with a
     radial scale length of 3.15 kpc. }
         \label{Fig_vdens}
   \end{figure}

\begin{figure}[!h]
   \centering
   \vspace{.4cm}
   \includegraphics[angle=-90,width=9cm]{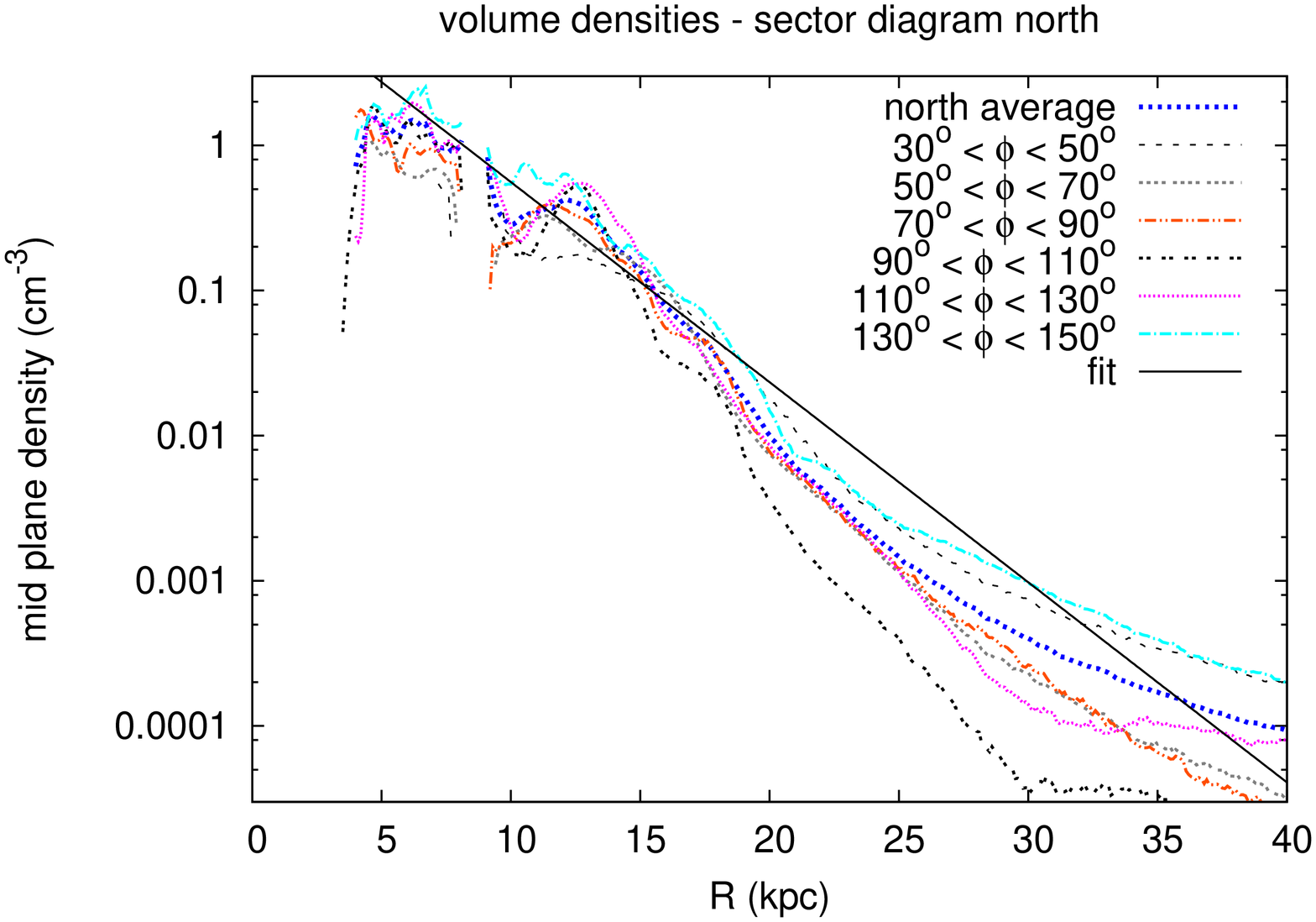}
   \includegraphics[angle=-90,width=9cm]{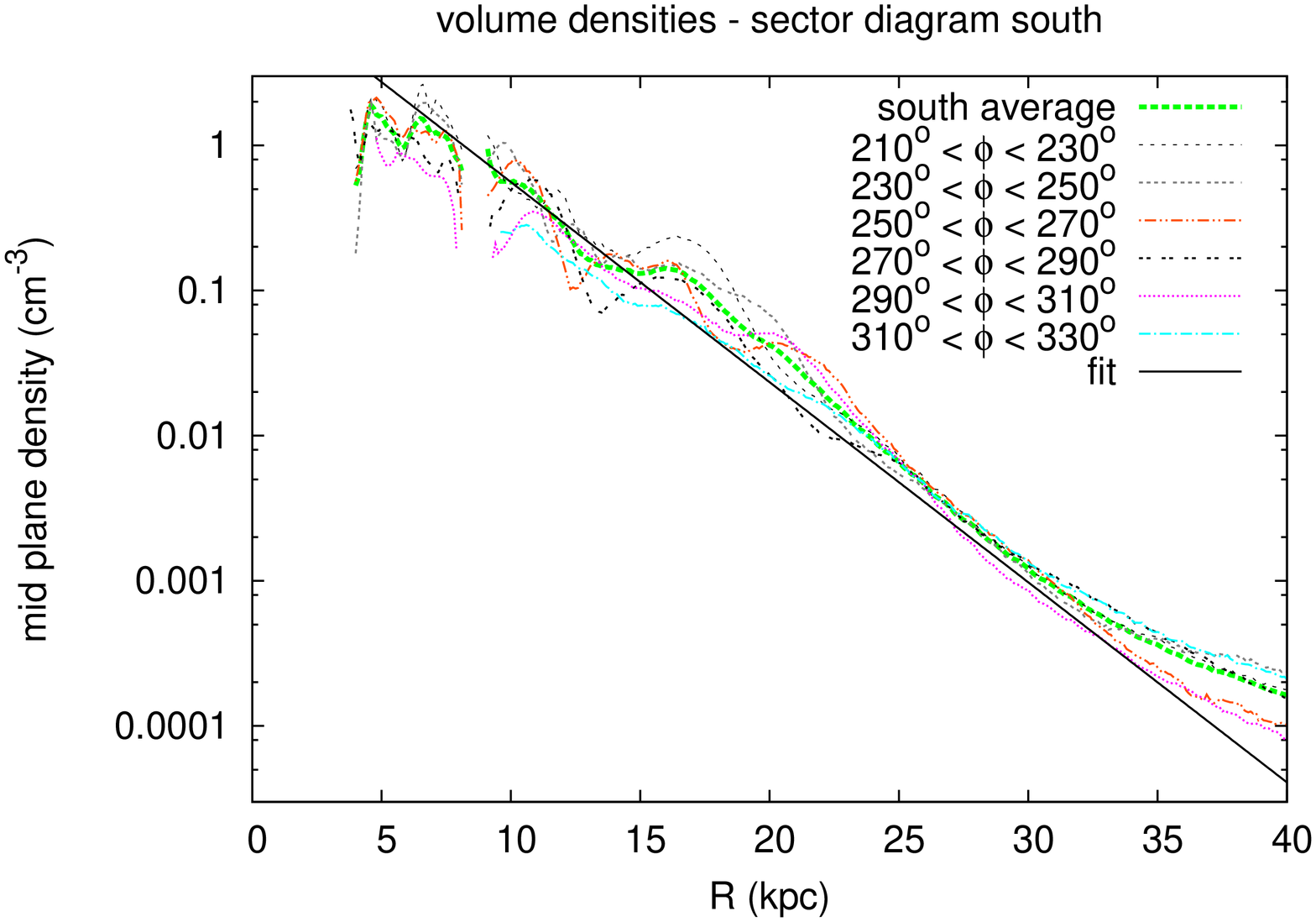}
   \caption{Systematic variations of the average \hi volume density
     distribution determined in $20\deg$ wide sectors for $30\deg < \phi <
     150\deg$ (north) and $210\deg < \phi < 330\deg$ (south). For comparison
     with Fig. \ref{Fig_vdens} the fit (black) and total averages for north
     and south (blue and green respectively) are given. }
         \label{Fig_vdens_scatter}
\end{figure}

Inside the solar circle the translation between distance and velocity is
ambiguous because there are two positions along the line of sight that have
the same velocity. These positions, however, may have different $z$ distances,
so their densities may differ accordingly. We modeled the expected flaring of
the \hi distribution and weighted the derived quantities correspondingly. Our
method is comparable to that of \citet{Nakanishi2003}, except that they use a
$sech^2$ approach \citep{Spitzer42} in place of our more detailed model. Our
model matches well with the observed flaring; for a verification of the model
we refer to \citet[][Fig. 14]{Kalberla2007}.

For a description of the global properties of the Galactic \hi disk we discuss
in the next sections the first three moments of the $n(R,z,\phi)$
distribution; the surface density, $\Sigma (R,\phi)$, the first moment
$z_0(R,\phi)$, representing the mid-plane, and the second moment $\sigma
(R,\phi)$, representing the scale height of the gas.  In addition we derived
characteristic mid-plane volume densities $n_0(R,\phi)$ for positions
$z_0(R,\phi)$, averaging over $z_0(R,\phi) \pm 0.6~ \sigma(R,\phi)$.

\section{The average radial surface density distribution}

Figure \ref{Fig_surf} displays average surface densities determined for the
total disk, but also for azimuth $ 30\deg < \phi < 150\deg$ (north) and $
210\deg < \phi < 360\deg$ (south) separately. There are some deviations
between north and south, we therefore derived the detailed surface density
distribution across the disk in individual sectors of $\Delta \phi = 20\deg$.
Figure \ref{Fig_sdens_scatter} displays the results separately for both parts
of the disk. Strong fluctuations, typically within a factor of two, with a
correlation length of a few kpc are common. Some of the modulations are
caused by a spiral structure \citep{Levine2006a,Levine2006b,Kalberla2007}.

Fig. \ref{Fig_sdens_scatter} confirms the general systematic differences
between north and south. Most striking, however, are two regions with
exceptional low fluctuations; in the north for $15 \la R \la 20$ kpc and in
the south for $25 \la R \la 30$ kpc. A comparison with the mass model and the
flaring data from \citet[][Figs. 18 \& 19]{Kalberla2007} indicates that these
regions are located at the outer edge of the dark matter ring structures that
were needed to explain the observed gaseous flaring. These structures have
different locations in the northern ($R \sim 13$ kpc) and southern part ($R
\sim 18.5$ kpc) of the Galactic disk but extend in azimuth over at least
120\deg.

Despite all the scatter, the global radial surface density distribution
appears to follow an exponential law over a broad radial range.  We
approximate the mean surface density distribution by $\Sigma(R) \sim s_0 \cdot
e^{-(R-R_{\sun})/{\rm R_s}} $ with $s_0 = 30$ \msun pc$^{-2}$ and ${\rm
  R_s} = 3.75$ kpc. The exponential disk approximation is reasonable for $
12.5 \la R \la 30 $ kpc. For $R \la 12.5 $ kpc the surface densities saturate
at approximately $\Sigma_{inner} \sim 10$ \msun pc$^{-2}$. This result is in
clear conflict with the contemporary knowledge from reviews \citep{DLARAA} and
text books; we reproduce in Fig. \ref{Fig_surf} the distribution of
\citet[][Fig. 9.19]{BM}. Dickey and Lockman give $\Sigma =$ 5 \msun pc$^{-2}$,
the average surface densities for $R \la 18$ kpc according to \citet{BM} is
even less. For $R \ga 18$ kpc our data agree with \citet[][Fig. 9.19]{BM};
a general scale error cannot explain the problem.

Discrepancies in Fig. \ref{Fig_surf} need a detailed discussion, fortunately
this is simplified greatly by a recent review
\citep[][Sect. 4.3]{Lockman2002a}. Figure 9.19 of
\citet{BM} dates back to \citet{Dame1993} who used data provided by
\citet{Lockman1988}. Lockman studied two cases, a surface density distribution
for a flat rotation curve and alternatively the case of a curve that is flat
for $R < $\rsun, rising at a rate $dv/dR = 2 $ \kms kpc $^{-1}$ up to 2
\rsuns and staying flat afterwards. In both cases it was assumed that the
scale-height of the \hi layer is constant. Some unspecified data were
``arbitrary scaled by a factor of 2'' by Dame to match other results. The case
of a rising rotation curve was then adopted by \citet[][Fig. 9.19]{BM} as
characteristic for the Milky Way.

A slightly different surface density distribution, also plotted in
Fig.\ref{Fig_surf}, was used by \citet{Wolfire2003}; we refer to their
detailed discussion. These authors match $\Sigma =$ 5 \msun pc$^{-2}$
according to \citet{DLARAA} with surface densities derived by
\citet{Wouterloot90}. Their curve rises for $R \ga 10$ kpc. At $ R \sim 12.5$
kpc it matches our average surface density distribution and at larger
distances our results agree well with \citet{Wouterloot90}; their data are
also plotted in Fig. \ref{Fig_surf}. In general we have for $R \ga 10$ kpc a
good agreement with
\citet{Henderson1982,Kulkarni1982,Diplas1991,Voskes1999,Levine2006a,Levine2006b}.

Attempting to explain the discrepancies for $R \la 10$ kpc, the determination
of \hi surface densities close to \rsun may be subject to serious
uncertainties, this was discussed in Sect. 2. In addition we found large
biases caused by the assumption that the \hi gas at $|z| > 0$ kpc is strictly
co-rotating with gas at the mid-plane \citep[][see Figs. 3 \& 5 for the
flaring]{Kalberla2007}. Such biases may depend on the methods used.  For a
co-rotation we derive a 10\% lower surface density for $R \la 12.5 $ kpc, at
the same time the exponential scale length increases from ${\rm R_s} = 3.75$
kpc to ${\rm R_s} = 4$ kpc. The slope is consistent with previous results
\citep{Henderson1982,Diplas1991,Voskes1999,Levine2006a,Levine2006b} but the
average surface density for $R \la 12.5$ kpc is still discrepant. We tested
the case of a rising rotation curve as proposed by \citet{deBoer2005} with a
peak rotational velocity of 250 \kms but found no significant changes of the
local surface density. This most probably is caused by the fact that locally
this rotation curve is flat. We question whether a rotation curve with a local
rate $dv/dR = 2 $ \kms kpc $^{-1}$ is realistic for the Milky Way.

\begin{figure}[!ht]
   \centering
   \includegraphics[angle=-90,width=9cm]{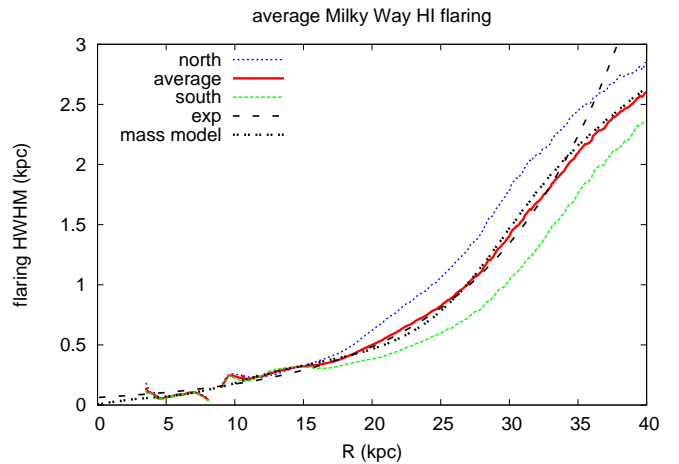}
   \caption{Derived mean flaring of the \hi gas (red). The dotted blue line
     holds for the northern, the dotted green line for the southern part of
     the Milky Way. For comparison an exponential fit for $R < 35$ kpc is
     plotted (long dashed), also the flaring according to the model by 
     \citet{Kalberla2007}. }
         \label{Fig_flare}
\end{figure}

\section{Average radial mid-plane volume densities}

Figure \ref{Fig_vdens} displays the average mid-plane volume density
distribution $n(R,z_0)$ for the total disk, for $30\deg < \phi < 150\deg$
(north), and $210\deg < \phi < 330\deg$ (south). Figure
\ref{Fig_vdens_scatter} shows how the volume densities depend on azimuth;
these plots reproduce for $R \ga 12.5$ kpc the same trends as seen before in
the surface densities (Fig. \ref{Fig_sdens_scatter}).

For $7 \la R \la 35 $ kpc the average mid-plane volume density follows 
approximately an exponential distribution, $n(R,z_0) \sim n_0 \cdot
e^{-(R-R_{\sun})/{\rm R_n}}$ with $n_0 = 0.9 $ cm$^{-3}$ and $R_n = 3.15 $
kpc. This line is drawn in Figs. \ref{Fig_vdens} and \ref{Fig_vdens_scatter}.

The surface density is essentially a product of mid-plane volume density and
the corresponding flaring. The fact that surface density as well as mid-plane
volume density can be approximated by an exponential function implies that
the flaring also needs to be described well by an exponential function. We
confirm this as the fit $h_R = h_0~e^{(R-R_{\sun})/R_0}$ kpc with $h_0 =
0.15$kpc and $R_0 = 9.8$ kpc for the half width at half maximum provides a
good approximation to the observed flaring \citep[][Fig. 6]{Kalberla2007} for
$5 \la R \la 35 $ kpc.  For $R \ga 35$ kpc, data and fit diverge
rapidly. Figure \ref{Fig_flare} re-plots for comparison the flaring derived in
\citet{Kalberla2007}.

The finding that surface density, mid-plane density, and flaring can be
represented by exponential functions does not reflect a general property of
galactic disks but is closely related to the rotation curve used, hence it
depends on the Galactic mass distribution.  In particular, the divergence of
the exponential fit from the observed flaring for $R \ga 35$ kpc is caused by
the shape of the gravitational potential which is oblate for small radii but
changes to prolate in the outer part. The transition region, the location
where we have a potential similar to the spherical case, is at $R \sim 35$ kpc
\citep[for a more detailed discussion see][Sects. 5.2.2 \&
5.2.3]{Kalberla2003}.  This radius defines the ``edge'' of the disk.

\section{The Galactic outskirts}

Common to the average surface density and the average mid-plane density is
that they both do not indicate a boundary of the Milky Way disk with a
truncation of the observable \hi distribution. Beyond a radius of $ \sim 35$
kpc the gas appears to have a more shallow distribution. We study this
behavior in more detail.

\begin{figure}[!t]
   \centering
   \includegraphics[angle=-90,width=9cm]{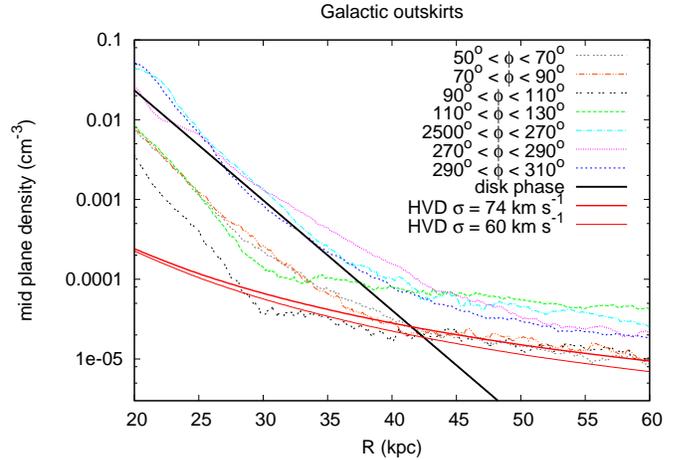}
   \caption{Average mid-plane volume densities for $20\deg$ wide
     radial sectors in the Milky Way. For comparison the fitted distribution
     of the HVD gas is given (red solid lines). }
         \label{Fig_vdens_out}
\end{figure}
\begin{figure}[!t]
   \centering
   \includegraphics[angle=-90,width=9cm]{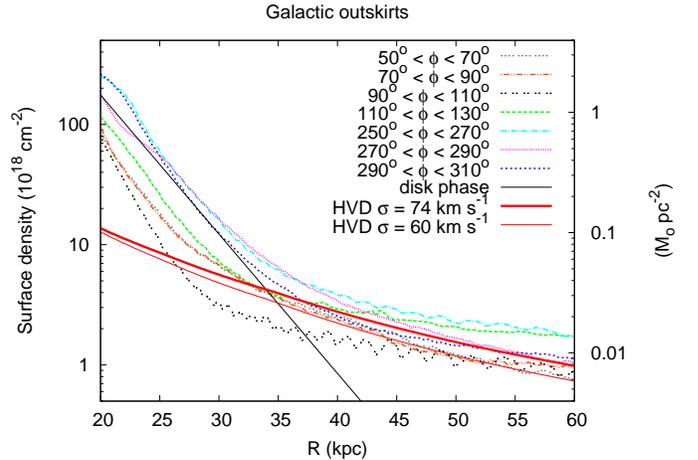}
\caption{Average mid-plane surface densities for $20\deg$ wide
     radial sectors in the Milky Way. For comparison the fitted distribution
     of the HVD gas is given (red solid lines). }
         \label{Fig_sdens_out}
\end{figure}

\subsection{The warp, scalloping} 

The Galactic warp was shown by \citet[][Figs. 16 \& 17]{Kalberla2007} to be
well defined for $R\la 40$ kpc. Three basic modes exist with slowly variable
phases.  The outermost part of the warp for $R \ga 30$ kpc contains higher
Fourier modes, thus the disk is scalloping.  We determine these modes by
subtracting the basic warp modes 0 to 2 from the derived mid-plane
positions. For the residuals we use a least square fitting algorithm to search
for oscillations with higher frequencies.  We determine simultaneously
amplitude, frequency, and phase. For various initial estimates in frequencies,
corresponding to modes 3 to 20, we demand that the fit stays stable in
frequency. As a result, we find oscillations with scalloping modes 10 to
20. The significance of these higher warp modes is low for $R \la 30$ kpc. The
associated power, defined by the squares of the amplitudes, is below 0.6\% of
the total warp power in this range. For $R \ga 30$ kpc we find a significant
contribution; the preferred modes are 10 and 13. The relative strength of the
scalloping increases linearly up to $\sim2.5$\% of the total warp power at $R
= 40 $ kpc. This is shown in Fig. \ref{Fig_wake} (magenta). The warp can be
traced further with the scalloping power increasing to larger distances but
the significance of the warp is highly questionable there.

Our analysis is complementary to \citet{Levine2006a} who used Morlet wavelet
transforms for an analysis of the warp out to $R = 30$ kpc. They concluded
that the m=10 and 15 scalloping modes are well above the noise, but localized;
this suggests that the scalloping in this range does not pervade the whole
disk, but only local regions. We applied a straightforward sine wave analysis
which is insensitive to local oscillations and found some coherent grouping of
higher mode oscillations, spreading out over several annuli, although
insignificant for fits at individual radii. Only for $R \ga 30 $ kpc does
scalloping become powerful enough to be significant in each case. This result
supports \citet{Saha2006} who concluded that scalloping at high modes has to
be restricted to the outermost part of a galaxy.

\begin{figure*}[!t]
   \centering
   \includegraphics[angle=-90,width=18cm]{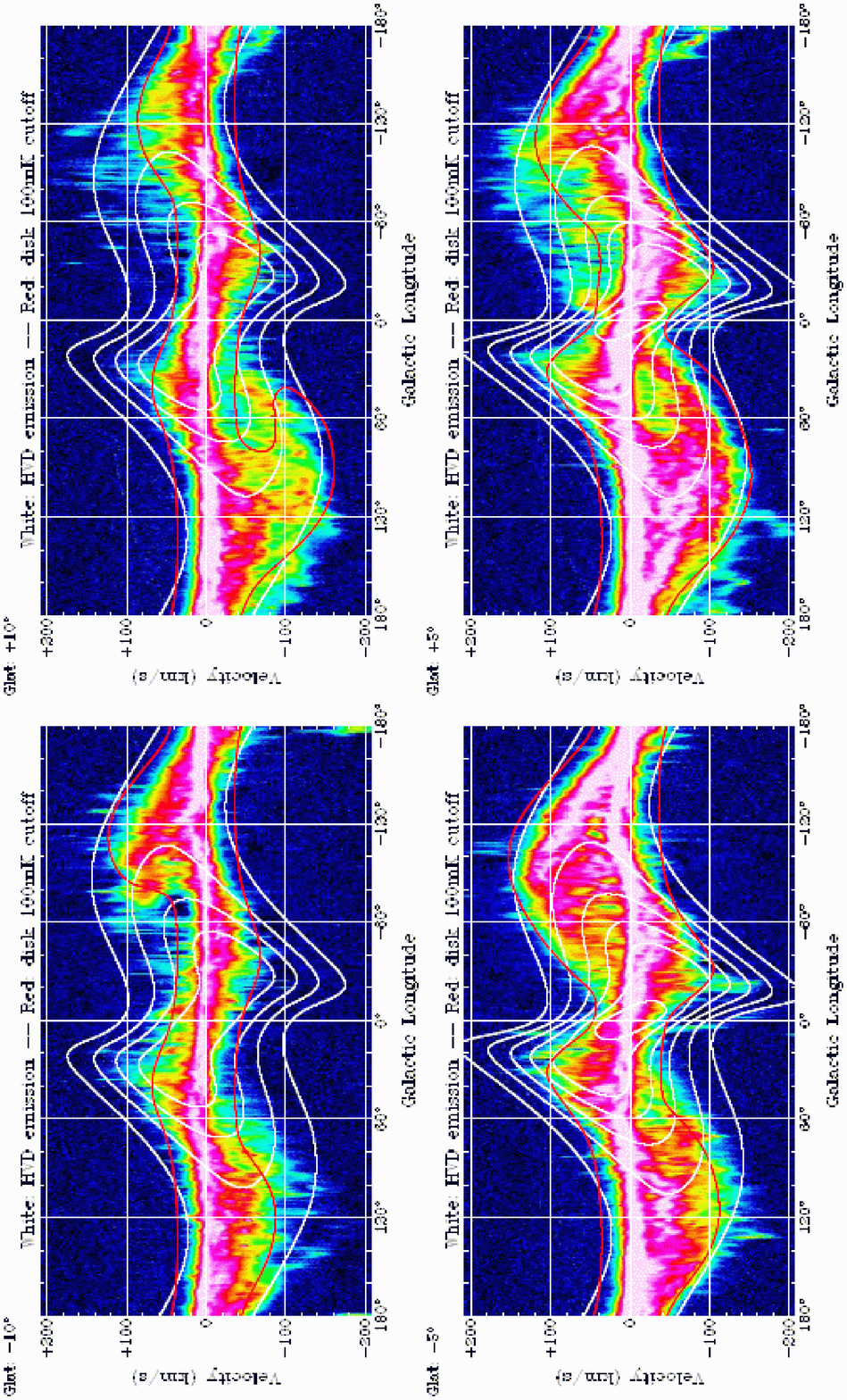}
   \caption{Position-velocity diagrams for the Milky Way \hi brightness
     termerature at latitudes -10\deg, -5\deg (left) and 10\deg, 5\deg
     (right). The white isophotes display HVD emission from the model (10\% to
     90\% in steps of 10\%), the red isophote marks an average disk emission
     of 100mK. A logarithmic transfer function was used for $T_B <30$K. }
         \label{Fig_lv_model}
\end{figure*}

\subsection{Mid-plane and surface densities}

Figure \ref{Fig_vdens_out} plots average mid-plane volume densities for
$20\deg$ wide sectors as used before. Comparing the distributions with the
exponential fit from Fig. \ref{Fig_vdens} we find a shallower decrease for $R
\ga 35$ kpc. Average surface densities displayed in Fig. \ref{Fig_sdens_out}
show the same trend. Both plots suggest that the \hi gas in the outskirts
differs significantly from the \hi gas in the Galactic disk, which was fitted
before by an exponential law.

Tracing the shallow \hi volume density distribution at large distances back to
the observed brightness temperature distribution we find that these belong to
weak emission profile wings. The wings cannot be explained by emission from
the cold or the warm neutral medium (CNM and WNM, respectively). Similar wings
have been found in the Leiden/Dwingeloo Survey \citep{Atlas1997} at high
latitudes and \citet[][Fig. 1]{Kalb98} argued that the wings can be traced to
low latitudes; this implies large distances. The LDS profile wings could be
fitted well by a pervasive emission component with a high velocity dispersion
(HVD) of $\sim 60$ \kms.

It is tempting to associate the shallow density distribution, visible in
Figs. \ref{Fig_vdens_out} and \ref{Fig_sdens_out}, with such an HVD
component. The proof, however, is not straightforward. The observed profile
wings must originate from a diffuse distribution of unresolved \hi clumps and
it is a priori not obvious how these wings translate into the corresponding
volume density distribution. We therefore modeled the \hi emission for an HVD
component with a radial scale length of $R_{HVD} = 7.5$ kpc
\citep{Kalberla2003}. Synthetic emission lines were calculated for a
dispersion of $\sigma_{HVD} = 60$ \kms and $\sigma_{HVD} = 74$ \kms
(differences will be discussed in Sect. 6.1).  Next we derived the surface
density and volume density distribution for the synthetic HVD distribution by
applying the same procedures as used before for the analysis of the LAB
observations. Figures \ref{Fig_vdens_out} and \ref{Fig_sdens_out} show the
result. The HVD component is weak in comparison to the disk emission at low
radii but dominates the \hi distribution at the outskirts for $R \ga 35$ kpc.

The assumption of a pervasive extended HVD \hi component agrees well with the
data (Fig. \ref{Fig_vdens_out} and \ref{Fig_sdens_out}). The HVD model with
$\sigma_{HVD} = 74$ \kms~ fits slightly better. As discussed before for the
disk gas, there are systematic differences between north and south. Where is
the HVD emission located in the observed $T_B(l,b,v)$ distribution?

In the Galactic outskirts the HVD gas becomes noticeable for $R \ga 35 $ kpc
and the range $ 50\deg \la \phi \la 130\deg $ and $ 250\deg \la \phi \la
310\deg $ is useful for its analysis. The translation between $T_B(l,b,v)$ and
$n(R,z,\phi)$, discussed in Sect. 2, is such that this gas must appear at high
negative velocities for longitudes $40\deg \la l \la 140\deg$ and at high
positive velocities for $ 310\deg \ga l \ga 220\deg$. Figure
\ref{Fig_lv_model} gives a few examples for the observed \hi brightness
temperature distribution, $l$--$v$ slices at latitudes $b = -10\deg, -5\deg,
5\deg$, and $b = 10\deg$.

Extra-planar \hi gas has a low volume filling factor. Correspondingly, a
patchy low level emission is expected in the region of interest. A second
property is the high velocity dispersion for this clumpy medium. This shows up
as extensions (stripes) in the $v$ direction. Figure \ref{Fig_lv_model} shows
a rather complicated pattern.  A part of the emission may originate from large
distances but the warp makes it difficult to determine whether individual
filaments belong to the disk (having a low velocity dispersion) or not. For a
clearer picture we calculated the expected average disk emission and plotted
the positions where this emission should reach a level of 100 mK (red
isophotes).  Then we include the average emission according to the
HVD model in Fig. \ref{Fig_sdens_out} (white isophotes).

In the inner Galaxy the HVD emission is clear in Fig.  \ref{Fig_lv_model} with
high velocities against the disk emission at longitudes $l \sim 25\deg$ and $l
\sim 335\deg$. The region at $l \sim 335\deg$ is currently studied by Alyson
Ford with the Parkes telescope.  The region $l \sim 25\deg$ was investigated in
detail by \citet{Lockman2002b,Lockman2004}. Numerous small clumps can be
traced with the GBT up to distances of several hundred pc above the
mid-plane. These clumps are usually unresolved in the LAB survey but
contribute to the observed surface densities.

Comparing the HVD model emission with observations, we find
for some regions significant deviations from the expected distribution. This
reflects the fact that HVCs and IVCs are grouped in complexes, while there are
also regions with little extra-planar \hi gas.

\subsection{Asymmetric gas distribution}

\begin{figure}[!ht]
   \centering
   \includegraphics[angle=-90,width=9cm]{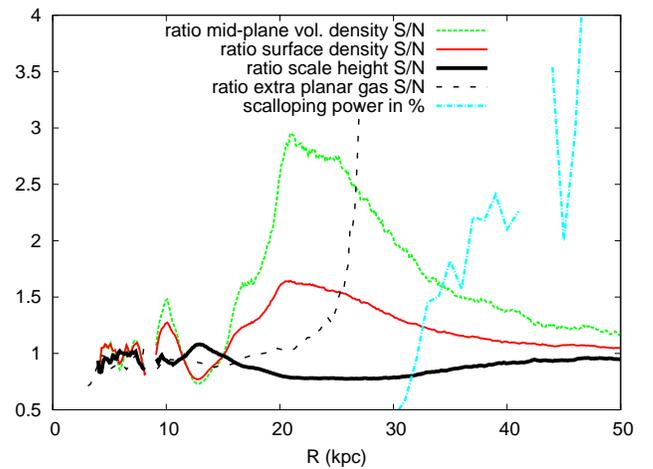}
   \caption{Asymmetry of mid-plane volume densities, surface densities, and
     scale heights characterized by their south/north ratios S/N. In addition
     the warp scalloping power in \% is given. }
         \label{Fig_wake}
\end{figure}

Figures \ref{Fig_surf} to 
\ref{Fig_sdens_out} show a pronounced asymmetry of the gas distribution for $
R \ga 20$ kpc.  For a more quantitative comparison we calculate the
south/north ratios (S/N) for mid-plane volume densities, surface densities and
for the scale heights. Fig. \ref{Fig_wake} demonstrates that the asymmetry is
strongest for mid-plane volume densities.

\begin{figure}[!t]
   \centering
   \includegraphics[angle=-90,width=9cm]{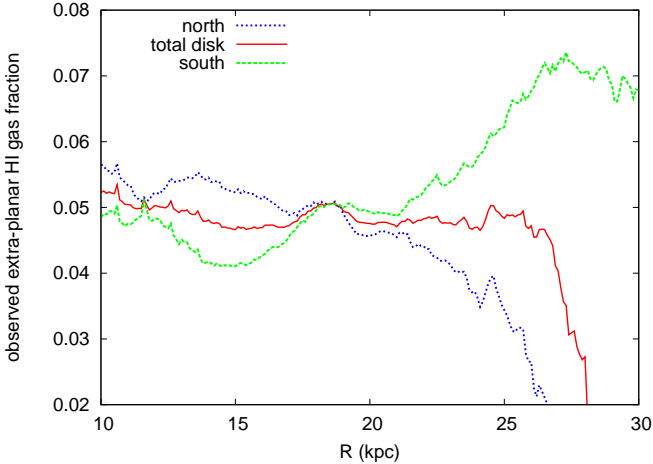}
   \caption{Fraction of extra-planar gas observed outside the disk for $ -15 <
     z < 20$ kpc. }
         \label{Fig_fract}
\end{figure}

We argued in Sect. 4 that the \hi gas in the Milky Way behaves on average as
expected for an exponential gaseous disk; the surface density is correlated
with the product of the mid-plane density and the scale height. This balance,
however, does not apply for the asymmetries visible in Fig. \ref{Fig_wake};
there is more gas in the south than in the north. The south/north asymmetry is
even more pronounced if we compare the extra-planar gas fraction for $R\ga 25$
kpc (Fig. \ref{Fig_fract}). The northern part of the extra-planar gas appears
abruptly torn away to the south for $R \ga 25$ kpc, in the direction of the
Magellanic System.

\section{Local gas high above the disk}
 
The distribution of planar and extra-planar gas is locally affected by
dynamical processes. Here we consider a steady state that is dictated by the
necessity of a global balance between gas pressure and gravitational
forces. We discuss first how a pressure balance can be achieved by a multi-phase
medium, next we use \hi line data to verify the derived model. We also compare
the extra-planar \hi gas in the Milky Way with the anomalous gas phase
observed in a few galaxies.

\subsection{Gravity and vertical gas pressure} 

\begin{figure}[!th]
   \centering
   \includegraphics[angle=-90,width=9cm]{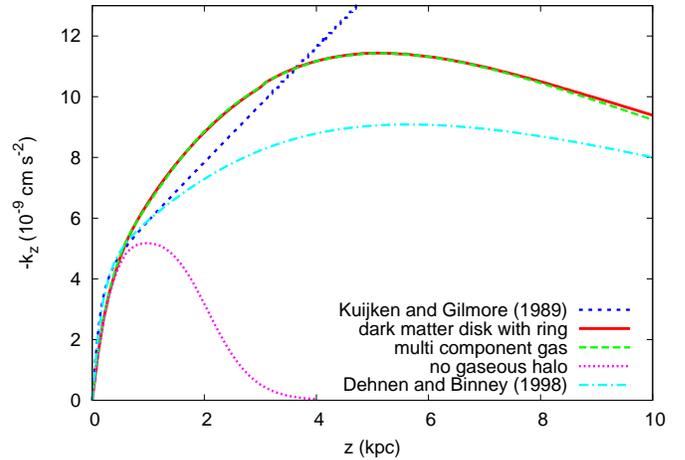}
   \caption{Gravitational acceleration $k_z$ at \rsuns = 8.5 kpc, (solid red
     line) and gas pressure balance $\rho ^{-1} dp/dz$ (dotted green). For
     comparison we plot $k_z$ according to \citet{KG89} (dashed, blue) and
     \citet[][Galactic Model 2]{Dehnen98}. The lower dotted line gives $\rho
     ^{-1} dp/dz$ for a multi-phase gas without halo components. }
         \label{Fig_kz}
\end{figure}

For a multi-phase medium, forces acting on the gas due to the pressure
gradient $\rho ^{-1} dp/dz$ need to balance the gravity $k_z$
\citep[e.g.][]{Boulares1990} and we may use this property to fit the
parameters of the gas distribution. We use the mass distribution provided by
\citet{Kalberla2007} to calculate $k_z$ at $R = 8.5$ kpc up to $z = 10$ kpc
(Fig.  \ref{Fig_kz}). The volume density of the extra-planar gas phase is
three orders of magnitude below the density of the other material in the
disk. Correspondingly, $k_z$ is unaffected by the halo gas (at the peak, $z
\sim 5$ kpc, the influence is $\la 0.3$\%); the extra-planar gas is a genuine
tracer of the Galactic potential.

We fit $\rho ^{-1} dp/dz$ for a multi-component gas to match $k_z$ and obtain
agreement with a mean error of 0.6\%. In Fig.  \ref{Fig_kz}, $\rho ^{-1}
dp/dz$ is over-plotted on  $k_z$; the gas pressure is corrected for a helium
abundance of 9\% of the hydrogen by number. Essentially, both curves are
identical, with differences hardly visible.  Table 1 lists the different phases,
densities $n_{fit}$ and velocity dispersions $\sigma_{fit}$ used to balance
the gas pressure gradient against the gravitational field. For comparison we
give $n_{obs}$ and $\sigma_{obs}$, derived previously by fitting the gas
distribution in the Solar neighborhood \citep[][Table 1]{Kalberla2003}. These
parameters agree well. The most significant deviation is found for the
velocity dispersion of the HVD component. This is a sensitive parameter in 
fitting the pressure gradient, but is less well determined when fitting observed
profile wings. In Sect. 5.2 we used both solutions for comparison.

The shape of $k_z$ for $|z| \la 1.1$ kpc is dominated by the stellar
distribution, see \citet{KG89}. We plot their $k_z$ relation for reference. At
large $z$ distances we find differences; the dark matter distribution
dominates the shape of the $k_z$ function and the gravitational forces become
model dependent. The model assumptions by \citet{KG89} lead to ever-rising
forces which are physically implausible. The forces have to decrease at large
$z$ distances \citep{Rohlfs1977}. Such a solution was provided by
\citet[][Galactic Model 2]{Dehnen98} and we plot their $k_z$ for
reference. This model is currently favored for ISM research
\citep[][Sect. 2.2]{Cox2005} and the downward bending of $k_z$ is interpreted
as a shaping provided by the non-planar geometry in the Dehnen \& Binney
model. In the model of \citet{Kalberla2007} the decrease of the forces at
large $z$ distances is caused by the mass associated with the disk.  The three
$k_z$ curves are discrepant for $k_z \ga 1.1$ kpc. Stars cannot be used to
decide between the models but we may use the gas distribution for this
purpose.

Attempting to fit the gas phase to match $k_z$ from \citet[][Galactic Model
2]{Dehnen98}, we need only 80\% of the observed density for the hot halo
phase, but 5.5 times the observed density for the neutral halo phase. This,
clearly, does not match the observations.  We conclude that the observed
extra-planar gas phase can be used to constrain dark matter models. Our
approach is consistent with the extra-planar gas as observed in the local
vicinity. However, some caution is needed. It is not expected that parameters
derived at the position of the Sun agree {\it exactly} with model parameters
for $R \sim 8.5$ kpc, since results are dependent on local influences,
predominantly caused by the local hot bubble. A hydrostatic model can predict
only {\it global} properties.
 
Current ISM models \citep[e.g.][Sect. 2.1]{Cox2005} do not include a hot or
neutral halo phase. For comparison, we plot $\rho ^{-1} dp/dz$ for the fitted
parameters from Table 1, but without a hot or neutral halo phase. This makes a
huge difference. For a global pressure balance at large $z$ distances a halo
gas phase is unambiguously needed. Preliminary evidence for such a phase from
interstellar absorption lines was given by
\citet{Muench1961}. \citet{Lockman91} have shown that more than two \hi
components are needed to explain the average local \hi emission. In the next
section we discuss the observational evidence for a multiphase \hi component
in more detail. We show \hi emission resulting from a neutral halo gas phase
and demonstrate how the observations compare with the model given in Table 1.

\begin{table}[h]
\caption[]{Local multi-component gas parameters used for
  Fig. \ref{Fig_kz} in comparison to parameters derived from
  observations \citep{Kalberla2003}. }
\begin{flushleft}
\begin{tabular}{lcccc}
\hline 
Component & $n_{fit}$ &  $n_{obs}$ & $\sigma_{fit}$ & $\sigma_{obs}$  \\
  & cm$^{-3}$  & cm$^{-3}$  & km s$^{-1}$ & km s$^{-1}$ \\
\hline
hot halo phase & .0018 & .0013 & 60.0 & 60.0  \\
neutral halo phase & .0014 & .0012 & 74.0 & 60.0 \\
\hline
DIG & .034 & .024  & 26.8 & 26.8  \\
WNM & 0.19 & 0.10 &  14.8 &  14.8  \\
CNM & 0.50 & 0.30 & 6.1  & 6.1  \\
\end{tabular}
\end{flushleft}
\end{table}

\subsection{The polar caps in \hi}

\begin{figure}[!th]
   \centering
   \includegraphics[angle=-90,width=9cm]{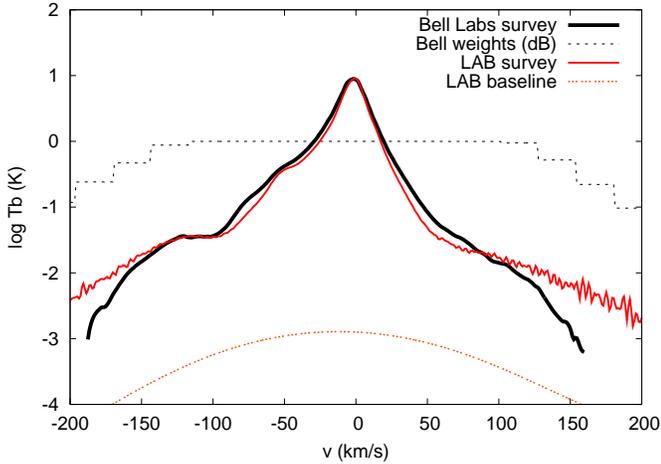}
\caption{Average \hi emission to logarithmic scale for latitudes $b >
  20\degr$. The data are from the Bell Labs telescope (thick 
  black line) and from the LAB survey (red). The step
  function shows the weights for each channel in dB. The thin red
  dotted line represents broad Gaussian components 
  indicating baseline problems. }
         \label{Fig_Bell}
\end{figure}

\begin{figure}[!ht]
   \centering
   \includegraphics[angle=-90,width=9cm]{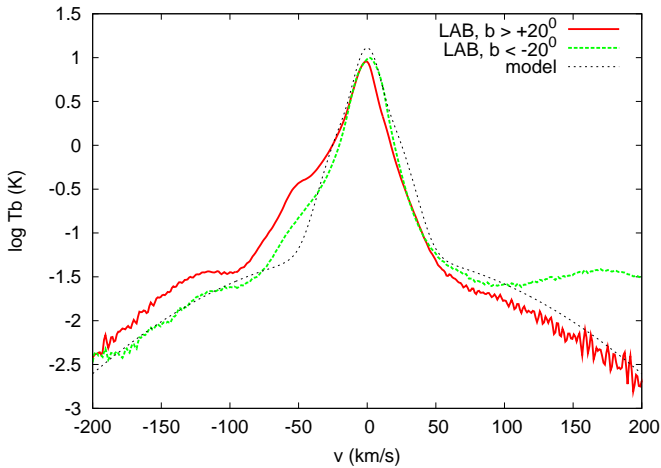}
\caption{Average \hi emission in logarithmic scale for latitudes $|b| >
  20\degr$ from the LAB survey. The red line is for the north polar
  cap, the dashed green line for the southern part. The thin dotted
  black line represents the model according to Table 1. }
         \label{Fig_LAB_20}
\end{figure}

Figure \ref{Fig_Bell} displays the average \hi emission measured at the north
polar cap for $ b > 20 \degr $. We use log($T_B$) to compress the huge dynamic
range of the \hi emission.  We compare data from the Bell labs survey
\citep[][Fig. 10]{Stark1992} with data from the LAB survey
\citep{Kalberla2005}. Both agree well, except for the wings at high 
velocities. Observational shortcomings cause the discrepancy. The bandwidth of
the Bell labs filter bank was too small to cover the necessary velocity range
completely; it needed to be shifted in velocity to obtain the main \hi
emission. The outer two channels of the filter bank were defined to be zero
for all spectra. The weights (in dB, dashed line) show how this affects the
mean Bell labs emission spectrum. As the weights decrease, the average Bell
labs profile becomes incomplete and biased.

The LAB survey may be influenced by systematic errors, too. The lower
red dotted curve in Fig. \ref{Fig_Bell} shows the mean of all broad LAB
Gaussians, most probably representing baseline defects in the LAB survey
\citep{Haud2006}. This is two orders of magnitude below the mean
emission; the weak profile wings are significant.

Figure \ref{Fig_LAB_20} compares the LAB data for the northern and southern
polar caps. We also plot the model distribution according to our best fit mass
model, using the fitted parameters for the \hi components from Table 1. Note
that the fit was to match $\rho ^{-1} dp/dz$ with $k_z$, but no fit was
applied to the plotted profile in Fig. \ref{Fig_LAB_20}, and to the wings
in Figs. \ref{Fig_vdens_out} \& \ref{Fig_sdens_out}.  All profiles in
Fig. \ref{Fig_LAB_20} show extended wings, but with some systematic
differences. The enhanced emission for $v_{lsr} \ga 100$ \kmss at the southern
polar cap is due to the Magellanic Stream. We find enhancements at $ v_{lsr}
\sim -60$ \kms, caused by distinct IVCs and at $ v_{lsr} \sim -120$ \kms,
caused by HVCs.

The broad wing underlying the CNM and WNM emission in Fig. \ref{Fig_LAB_20}
is caused by extra-planar gas, the HVD component. Numerous clumps, filaments,
and spurs, located above the CNM and WNM disk, contribute to a signal which
can be well represented by a weak component with a large velocity dispersion.
We conclude that the {\it local} observed extra-planar \hi gas distribution,
forced to be on average in pressure balance $k_z \sim \rho ^{-1} dp/dz$ 
(Fig. \ref{Fig_kz}), is consistent with the gas component that was proposed
in Sect. 5.2 to explain the gas distribution at the {\it outskirts} of the
Galaxy. Systematic differences are caused by a warped and asymmetric gas
distribution. Gas in the southern part of the outer Milky Way appears to be
enhanced.  This is consistent with the enhanced positive velocity wing in
Fig. \ref{Fig_LAB_20} for $b < 20\deg$. The negative profile wing is
enhanced for $b > 20\deg$ due to the warp that bends up to the
north.

\subsection{Extra-planar gas fraction} 

During our analysis we distinguished disk emission and extra-planar gas. For a
safe determination of the first moments of the stationary disk gas we masked
and excluded outliers of the \hi volume density distribution which are most
probably not associated with the disk. This part of the $n(R,Z,\phi)$
distribution has radial velocities that are incompatible with a circular
rotation of the disk gas around the Galactic center. Individual features show
up as sheets and filaments in the $n(R,Z,\phi)$ distribution (see
Fig. \ref{Fig_R_cuts} or \citet[][Sects. 2.2 \& 3 ]{Kalberla2007}. The
corresponding clouds, visible in the $T_B(l,b,v)$ distribution
(Fig. \ref{Fig_lv_model}), are usually classified as HVCs and IVCs. In general
it is assumed that IVCs are located in the lower Galactic halo while HVCs are
at larger distances \citep{Albert2004}.

For a determination of the extra-planar gas fraction we use a range where the
properties of the Galactic disk are well defined, $10 \la R \la 30$ kpc. On
average we find there direct evidence that about 5\% of the \hi is not
associated with the disk (Fig. \ref{Fig_fract}).  This estimate is independent
of the mass model used. Using parameters from Table 1, integrating for $R \la
50$ kpc, we derive that the HVD component amounts in total to 10\% of the gas.
This discrepancy is explained by the fact that most of the HVD gas phase in
the Milky Way cannot be distinguished from the CNM/WNM gas phase. Only gas at
the wings of the HVD distribution is counted and we estimate that about 50\% of
this gas remains undetected by our filter and revise the extra-planar gas
fraction to be 10\%, consistent with the HVD model.

\subsection{Anomalous gas in the Milky Way, an external viewpoint }

\begin{figure}[!ht]
   \centering
   \includegraphics[angle=-90,width=9cm]{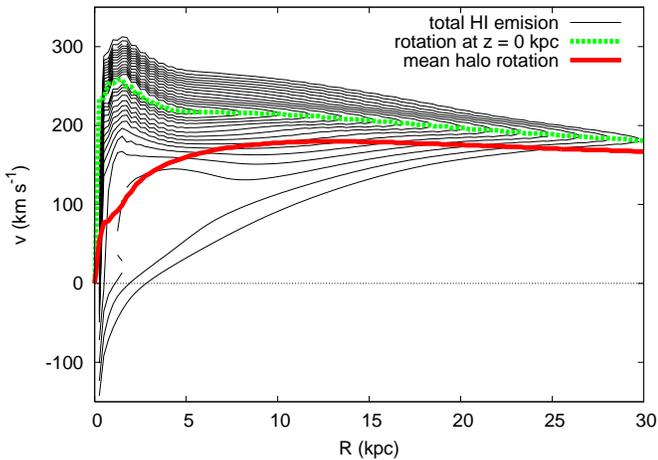}
   \caption{Position velocity diagram along the major axis, observable for a
     galaxy like the Milky Way with an inclination of $80\degr$.  The thin
     contours represent normalized column densities in steps of 2 dB; the
     lowest contour is at -30 dB. The green dashed lines gives the rotation
     curve at $z =0$ kpc, the solid red line the mean rotation for the \hi in
     the halo.  }
         \label{fig-beard_80}
\end{figure}

To compare the Milky Way with sensitive observations of other galaxies, we
take the viewpoint of an external observer. The \hi emission is synthesized
accordingly. We use a flat disk without warp at an inclination of $80\degr$
and calculate the total emission along the line of sight.  The contours in
Fig. \ref{fig-beard_80} represent observable column densities, normalized and
plotted in steps of 2 dB. We consider observations with a limited dynamic
range of 30 dB; the lowest contour is at -30 dB.

The simulated position velocity diagram shows that the velocity at the peak of
the line emission is identical with the rotation curve from the model. The
average observable rotation velocity for the halo gas phase alone, determined
from the peak emission of the halo gas phase lags behind. The total column
density distribution is symmetric at large column densities, but as a
consequence of the lagging halo some asymmetries appear at low
levels. Projection effects cause extended regions with \hi emission at lower
velocities. Such features at levels below -20 dB were first observed in NGC
891 \citep{Swaters1997}, the so-called beards \citep{Fraternali2001}. Close to
the center, the beards extend to forbidden velocities, mimicking an apparent
counter-rotation.

Beards and forbidden velocities in Fig. \ref{fig-beard_80} have the same
origin. The line of sight passes through various layers of extra-planar gas
that rotate slower than the disk. The lag is approximately constant for
$z/h_z(R)$ with $z$ distances normalized to scale heights
\citep[][Fig. 11]{Kalberla2003}. For small radii, the densest part of the
thick extra-planar gas disk close to the center becomes most prominent. The
lag is most pronounced there.

Our model takes both the flaring and lagging halo into account. Both are
necessary to explain the observations. The beards in Fig. \ref{fig-beard_80}
originate from the same \hi halo gas phase that causes the extended wings in
Figs. \ref{Fig_Bell} \& \ref{Fig_LAB_20}.  At the end of this subsection we
point out that an external observer, with a telescope that is limited in
dynamic range to 30 dB, would see the Galactic \hi disk out to $R \sim 30$
kpc only.  At low levels it is far easier to find extended emission
perpendicular to the disk as shown recently by \citet{Oosterloo2007} in the
case of NGC 891.  Reaching the outskirts at distances of $R \sim 50$ kpc, is a
demanding task and needs a dynamic range of 40 dB or better. The HVD
component at such radii is essentially invisible.

\section{Summary and discussion }

We used the LAB survey to derive the \hi volume density distribution in the
Milky Way. The $T_B(l,b,v)$ to $n(R,z,\phi)$ conversion depends on the
rotation curve; we have chosen an almost flat curve with lagging rotation for
gas above the mid-plane \citep{Kalberla2007}. The corresponding mass model implies
that some of the dark matter is associated with a thick exponential disk. 

Our aim was to derive global properties for the \hi gas distribution in the
Milky Way. We calculated the first moments to determine surface densities, 
mid-plane positions, associated mid-plane densities, and scale-heights. Planar
and extra-planar gas was distinguished by applying adaptive filters. We
discuss mid-plane positions first.

\subsection{The warp} 

The Galactic warp can be traced from $R \sim 12.5$ kpc out to $R \sim 45$ kpc
but the coherent structure is lost for $R \ga 40$ kpc. The scalloping power,
defined by the squares of the amplitudes from a sine wave fit, is below 0.6\%
in the inner part of the disk. For $30 \la R \la 45 $ kpc we observe a linear
increase up to 3\% of the total warp power.

\subsection{The exponential \hi disk}  

We derive average surface densities and, taking the warp into account, average
volume densities at mid-plane. In both cases we find significant fluctuations
and asymmetries between north and south which are correlated. Surface
densities and mid-plane volume densities are affected by spiral structure, but
differences between north and south need to be explained by an asymmetric disk
\citep{Levine2006a,Levine2006b,Kalberla2007}.
 
Average surface densities, mid-plane volume densities, and scale heights can
be approximated over a broad radial range by exponential relations. For the
surface density we find a scale length of 3.75 kpc within $ 12.5 \la R \la 30
$ kpc.  Mid-plane volume densities are well approximated with a scale length
of 3.15 kpc for $ 7 \la R \la 35 $ kpc. Both relations imply that the scale
heights also can be described well by an exponential relation, in this case
valid for $ 5 \la R \la 35 $ kpc.

For $R \la 12.5 $ kpc we find approximately constant surface densities
($\Sigma_{inner} \sim 10$ \msun pc$^{-2}$). This is also the inner part of the
\hi disk which is unaffected by the warp. 

\subsection{The Galactic outskirts} 

The exponential disk approximation for the disk gas breaks down in all
respects at radial distances $ R \ga 35$ kpc. In the outer part we find a
shallower distribution which can be approximated by a gas phase with a radial
scale length of 7.5 kpc and a high velocity dispersion of 74 \kms.

Our results are apparently in disagreement with \citet{Maloney1993}, who found
a sharp truncation of the neutral hydrogen distribution for the outer disk \hi
gas in NGC 3198 at a surface density level $ 10^{18} \la N_H \la 10^{19} $
cm$^{-2}$. This limit is attributed to the extragalactic radiation field. To
study the origin of the HVD component at large distances we examined the data.
At a distance of 60 kpc the telescope beam covers a typical area of 600 pc in
diameter. Any emission within this area is smoothed out completely. Inspecting
the data cube visually, we find regions with peak densities of $n(R) \sim
10^{-3}$ cm$^{-3}$, surrounded by less dense envelopes. These unresolved
individual condensations may be well above the predicted cut-off limit.

Maloney's interpretation was questioned recently by \citet{Oosterloo2007} who
found that a deeper mapping of NGC891 led predominantly to the detection of
unknown low level emission in the vertical direction, while the radial limits
remained. Such an asymmetry is unexpected for an isotropic radiation field.
Our analysis shows the same trend. At the outskirts of the disk the HVD
component becomes most extended in $z$ direction. Mid plane volume densities
are dominated by the HVD component for $R \ga 40$ kpc, while surface densities
(integrated in $z$ direction) show the same effect for $R \ga 35$ kpc (compare
Figs. \ref{Fig_vdens_out} and \ref{Fig_sdens_out}).

\subsection{Polar caps and extra-planar gas } 

The \hi gas distribution, averaged over all latitudes $|b| > 20\deg$, 
probing approximately the local volume within a radius of 1 kpc, has broad
profile wings. These wings may be affected by stray radiation of the telescope,
but for the Bell Labs and the LAB survey they are not explainable by 
instrumental problems. The same HVD component with a dispersion of 74 \kms,
which was used to fit the gas distribution in the outskirts, is useful also to
explain this local component. A multi-phase medium containing this HVD
component is found to balance gravity by its pressure gradient, $k_z \sim \rho
^{-1} dp/dz$ for the mass model proposed by \citet{Kalberla2007}.

\subsection{Clumps, fast and at large $z$ distances } 

Evidence for a population of fast gas clouds with a velocity dispersion of
$\sigma \sim 35$ \kms~ was found by
\citet{Radhakrishnan1980,Kulkarni1985,Lockman91, Mohan2004}. \citet{Kalb98},
analyzing the Leiden/Dwingeloo Survey, proposed a more extreme dispersion of
$\sigma \sim 60$ \kms~ which is modified here to $\sigma \sim 74$ \kms.  Most
remarkable, however, is the early prediction by \citet{Pikelner1958} who
proposed a neutral halo gas phase with $\sigma \sim 70$ \kms.

The HVD component analyzed here as a diffuse emission component with a
dispersion of $\sigma \sim 74$ \kms, is patchy. It is made up of a large
number of filaments and small \hi clumps. These stand out from the normal disk
emission only if they reach large $z$ distances or if they have high forbidden
velocities. Evidence for such clumps in the inner Galaxy was given by
\citet{Lockman2002b}. \citet{Stanimirovic2006} found similar clumps in the
anti-center direction. Clouds with high forbidden velocities in the inner
Galaxy were detected by \citet{Stil2006} in the VLA Galactic Plane
Survey. These authors estimate that this population of fast clouds has a
radial scale length of 2.8 to 8 kpc, consistent with the scale length of 7.5
kpc derived by us.

\subsection{Extra-planar gas fraction} 

For our analysis we used adaptive filters to distinguish the stationary disk
component from extra-planar gas. Over the main body of the disk ($10 \la R \la
30$ kpc) we measure that on average 5\% of the gas is outside the disk
(Fig. \ref{Fig_fract}). This is a lower limit only. Assuming that the
extra-planar gas can be characterized as an HVD component implies that a
significant fraction of this phase is hidden from direct observation since it
cannot be distinguished from disk gas. For our analysis we estimate that only
50\% of the LVD gas can be discriminated, resulting in an extra-planar gas
fraction of 10\%. This is consistent with a hydrostatic model (Sect. 6.1),
observations at the polar caps (Sect. 6.2) and also with the gas distribution
at the Galactic outskirts (Sect. 5.2).  We conclude that {\it all over the
  Galactic \hi disk} the ``normal'' phase, the CNM and WNM, is associated by
an ``anomalous'' gas phase, described here as an HVD phase. For the inner
Galaxy ($R \la 10$ kpc) we observe an increase of the extra-planar gas
fraction of a few percent. This might be explained by fountain activities but,
as discussed in Sect. 2, estimates at such distances are highly uncertain.

In this light, the general features of the extra-planar \hi gas in NGC 891 and
NGC 2403 appear to be comparable (within a factor of 2) to what we see in the
Milky Way. This is confirmed by projecting the \hi gas distribution of the
Milky Way to large distances. Simulating position-velocity diagrams, the HVD
component causes wings and beards close to the center. The gaseous outskirts,
however, would hardly be observable. The telescope sensitivity needs to be
improved by a factor of 10 to observe this part, which is feasible with the
SKA.


\subsection{HVC and IVC gas} 

HVCs and IVCs are usually classified from observations according to their
velocities. One can distinguish between coherent complexes and isolated
compact objects. Their origin may be from fountain flow or accretion. Also,
metalicities are important but here we discuss \hi properties only.

For our analysis of the global properties of the \hi distribution we have
chosen different criteria. The primary goal was to derive the properties of
the main body, the \hi disk. This has a well defined dynamically cool gas
distribution. There are two phases, CNM and WNM, both with velocity
dispersions perpendicular to the disk that are low compared to the rotation
velocity. The gas disk is flaring but the most prominent inner part of the \hi
disk is highly flattened.

However, as proposed by \citet{Spitzer1956} and much later established by
X-ray observations of the predominantly hot ISM of elliptical galaxies, the
hot Galactic ISM does not have to be so flat. The extra-planar \hi is embedded
in the hot component. This arises the question of whether there might be a
quasi-static system near equilibrium as modeled by Spitzer or whether the
Milky Way halo is strongly affected by steadily changing conditions, e.g. due
to mass inflow or star bursts.

The star-formation-rate (SFR) in the solar neighborhood is remarkably constant
over the lifetime of the disk, infall of material appears also to be
approximately constant at one-half of the SFR \citep{Twarog1980}. Ongoing
accretion may be derived from HVC complexes and also from the Magellanic
Stream but it is unclear whether irregularities in accretion rate and outflows
are strong enough to affect the global shape of the Galactic halo
significantly. Little is known about the impact of star bursts on the
structure and dynamics of the Milky Way halo. Supershells and chimneys may
reach $z$ distances of several kpc but appear not to be able to shape a
significant fraction of the halo \citep{McClure-Griffiths2006}.

As a halo model therefore we have chosen the simple case of a quasi-static
system that can be approximated numerically by a hydrostatic distribution. We
are interested in {\it global} properties of the gas distribution, not in
details that are more important for individual components. A global model
makes sense when averaging the \hi emission over several kpc as we obtain a
self-consistent solution. As discussed in Sects. 5 \& 6, the same model
parameters apply to both polar caps and to the sectors $ 50\deg \la \phi \la
130\deg $ and $ 250\deg \la \phi \la 310\deg $ at $40 \ga R \ga 60$ kpc. These
are the only regions available for an unambiguous separation of planar and
extra-planar gas in the Milky Way.

Figure \ref{Fig_lv_model}, a $l$-$v$ plot of measured brightness temperatures,
displays numerous IVCs and HVCs. These features appear partly to be organized
in complexes but are in general close to the disk. Projection effects cause a
complicated picture, in particular velocity crowding in some parts of the sky
makes it difficult to judge whether a particular feature belongs to the disk
or not. Deviation velocities between planar and extra-planar gas components
depend strongly on position.

The density distribution $n(R,z,\phi)$ in Fig. \ref{Fig_R_cuts} shows a
clearer picture. Extra-planar gas is offset from the Galactic disk but most of
it is clearly associated with the disk. Again we find large scale fluctuations
but there is a general trend that the extra-planar gas distribution increases
with decreasing mid-plane distance. This situation appears best to be described
as a ``thick disk'' having twice the radial scale length of the CNM/WNM gas
distribution. We use the model of a turbulent HVD distribution. Local
enhancements, or complexes, are considered as statistical fluctuations; they
become less dominant if one averages over larger areas, this is a typical
property of a turbulent medium. Figure \ref{Fig_LAB_20}, showing average
emission of the polar caps at latitudes $|b| > 20\deg$ also includes IVCs,
HVCs, and the Magellanic stream. These local phenomena show up as minor
perturbations.

Comparing properties of the HVD component with global properties of the HVC
gas we find some interesting agreements. HVCs have a multi-phase structure and
are most probably confined by an external hot halo phase
\citep{Kalberla2006}. HVCs have specific turbulent energy densities that are
an order of magnitude higher than that of comparable clumps in the Galactic
disk.  The same conditions hold for the HVD component. Most surprising,
however, is that the integrated flux distribution for HVCs as function of
velocity \citep[][Fig. 5]{Wakker1991} is a continuation of the gas
distribution in the broad wings of the HVD component to high velocities
(Fig. \ref{Fig_LAB_20}).

The HVD component is coupled to the disk but it is anomalous in the sense that
it represents a highly turbulent clumpy phase, best described as an extended
very thick disk. At the position of the Sun we determine an exponential scale
height of 3.9 kpc. The HVD component can be best characterized as an
extra-planar gas component; it expands the phase space occupied by the \hi
disk predominantly perpendicular to the disk. IVCs are located in the lower
Galactic halo \citep{Albert2004}. Their velocities as well as known $z$
distances place them at the less distant part of the HVD phase
space. Distances for HVCs are less well known but according to their
velocities at least some of them may belong to the outer part of the HVD phase
space. The HVC velocity limit is $|v_{lsr}| = 465$ \kms, roughly coincident
with the escape velocity of the Milky Way.

Individual extra-planar gas layers are extremely patchy and lag behind the
underlying rotating disk. These layers are affected by flaring and the lag
depends on the distance $z/h_z(R)$, normalized to the disk scale height. The
anomalous gas phase is rather extended. We could trace it up to $R \sim 60$
kpc.

The gravitational potential at the outskirts of the Milky Way is very shallow.
The rotation of anomalous gas for $R \ga 35$ kpc is therefore essentially
unaffected by flaring. The fact that we observe an extended HVD gas phase at
$40 \la R \la 60$ kpc implies accretion by a ``drizzle'' of low mass
fragments. This gas cannot originate from fountain flows. Fountain events far
outside $R_{25}$ are highly improbable as the radial scale lengths for stars
and HVD gas differ significantly.

A continuous flow of material in galaxies from the disk into the halo was
studied by \citet{Collins2002,Fraternali2006,Heald2006}. Their models can
explain the vertical gas distribution but fail to reproduce the apparent
counter rotation and vertical velocity gradients. They also produce a general
outflow, contrary to observations.

\subsection{The lopsided \hi distribution} 

Our aim was to describe global features of the \hi distribution that may be
representative of a steady state. Calculating radial distribution functions we
found in all cases significant systematic differences between the northern and
southern part of the Milky Way. The system is asymmetrical. Midplane volume
densities and surface densities are most discrepant at $R \sim 20$ kpc.
Considering the \hi flaring we found a systematic modulation that may be
explainable by a misaligned gravitational field originating from a misaligned
dark matter disk (Fig. \ref{Fig_wake}). The flaring ratio peaks at $R \sim 25$
kpc are in excellent agreement with \citet{Weinberg1998} who predicted that
the interaction between Milky Way and Magellanic system causes a dark matter
wake at this position.

\subsection{Rotation curve and mass distribution} 

Our results depend on the rotation curve, hence on the mass model. We tested
several different possibilities and summarize briefly how much our conclusions
are affected by the mass model used \citep{Kalberla2007}.

The global surface density distribution, mid plane densities and also flaring
depend weakly on the rotation curve
\citep{Voskes1999,Nakanishi2003,Levine2006a,Levine2006b} but our results are
not altered significantly as long as a nearly flat rotation curve is used.
The radial exponential scale length of 3.75 kpc derived by us changes to 4 kpc
if one uses a constant rotation with a circular velocity of 220 \kms. Our
conclusion that surface densities, mid plane densities and flaring can be
approximated by exponential relations remains valid, although the correlation
degrades somewhat if we use alternative rotation curves.

Deviations from axisymmetry depend on non-circular terms in the rotation
curve.  We used the first order epicyclic streamline correction by
\citet[][Eqns. 1\&2]{Levine2006a} which is a great improvement in overcoming
discontinuities at large radial distances. This correction is empirical but
the assumption that surface densities should not have azimuthal
discontinuities remains valid for all mass models. The correction is
consistent with an asymmetrical mass distribution and the results do not
depend significantly on the details of the mass model. The same holds for
asymmetries in the \hi distribution.

The average \hi emission at the polar caps (Sect 6.2) has broad high velocity
wings. This is known from observtions. Parameters for the underlying HVD
component are completely independent of any mass model. The parameter fitting,
assuming halo gas in global hydrostatic equilibrium with the gravitational
field, depends strongly on the mass model. A dependence on the mass model
exists also for HVD gas at the outskirts of the Milky Way. It is the shape of
the outer rotation curve that determines the distance of this component.

We argued that the thin \hi disk, defined by the CNM \& WNM gas, is
associated with a thick extended HVD gas disk. This model is self-consistent for
the mass model used but it necessarily has to break down for significant
changes of the mass model or the outer rotation curve.


\begin{acknowledgements} 
  This project was supported by Deutsche Forschungsgemeinschaft, grant
  KA1265/5-1. We thank J. Kerp and F. Kenn for discussions and a careful
  reading of the manuscript and the anonymous referee for constructive
  criticism.
\end{acknowledgements}


\begin{thebibliography}{}
\bibitem[Albert \& Danly(2004)]{Albert2004} Albert, C.~E., \& 
Danly, L.\ 2004, ASSL Vol.~312: High Velocity Clouds, 73 

\bibitem[Bajaja et al.(2005)]{Bajaja2005} Bajaja, E., Arnal, 
E.~M., Larrarte, J.~J., Morras, R., P{\"o}ppel, W.~G.~L., \& Kalberla, 
P.~M.~W.\ 2005, \aap, 440, 767 

\bibitem[Binney \& Merrifield(1998)]{BM} Binney, J., \&
Merrifield, M.\ 1998, Princeton, NJ : Princeton
University Press, 1998.

\bibitem[Boulares \& Cox(1990)]{Boulares1990} Boulares, A., \& Cox, 
D.~P.\ 1990, \apj, 365, 544 


\bibitem[Collins et al.(2002)]{Collins2002} Collins, J.~A., 
Benjamin, R.~A., \& Rand, R.~J.\ 2002, Bulletin of the American 
Astronomical Society, 34, 709 

\bibitem[Cox(2005)]{Cox2005} Cox, D.~P.\ 2005, \araa, 43, 337 

\bibitem[Dame(1993)]{Dame1993} Dame, T.~M.\ 1993, Back to the 
Galaxy, 278, 267 

\bibitem[de Avillez(2000)]{deAvillez2000} de Avillez, M.~A.\ 2000, 
\mnras, 315, 479 

\bibitem[de Boer et al.(2005)]{deBoer2005} de Boer, W., Sander, 
C., Zhukov, V., Gladyshev, A.~V., \& Kazakov, D.~I.\ 2005, \aap, 444, 51 

\bibitem[Dehnen \& Binney(1998)]{Dehnen98} Dehnen, W.~\& Binney, 
J.\ 1998, \mnras, 294, 429 

\bibitem[Dickey \& Lockman(1990)]{DLARAA} Dickey, J.\ M.\ \& Lockman, F.\ 
  J.\ 1990, \araa, 28, 215


\bibitem[Diplas \& Savage(1991)]{Diplas1991} Diplas, A., \& 
Savage, B.~D.\ 1991, \apj, 377, 126 


\bibitem[Fraternali et al.(2001)]{Fraternali2001} Fraternali, F., 
Oosterloo, T., Sancisi, R., \& van Moorsel, G.\ 2001, \apjl, 562, L47 

\bibitem[Fraternali et al.(2004)]{Fraternali2004IAUS} Fraternali, F., 
Oosterloo, T., Boomsma, R., Swaters, R., \& Sancisi, R.\ 2004, IAU 
Symposium, 217, 136 

\bibitem[Fraternali \& Binney(2006)]{Fraternali2006} Fraternali, F., \& Binney, J.~J.\ 2006, \mnras, 366, 449 

\bibitem[Hartmann \& Burton(1997)]{Atlas1997} Hartmann, D.~\& 
Burton, W.~B.\ 1997, Atlas of Galactic Neutral Hydrogen 
(Cambridge: Cambridge University Press)


\bibitem[Haud \& Kalberla(2006)]{Haud2006}Haud,~U.\, \& Kalberla,
P.~M.~W.  2006, Baltic Astronomy, 15, 413 

\bibitem[Heald et al.(2006a)]{Heald2006} Heald, G.~H., Rand, 
R.~J., Benjamin, R.~A., Collins, J.~A., \& Bland-Hawthorn, J.\ 2006, \apj, 
636, 181 

\bibitem[Henderson et al.(1982)]{Henderson1982} Henderson, A.~P., 
Jackson, P.~D., \& Kerr, F.~J.\ 1982, \apj, 263, 116 

\bibitem[Kalberla et al.(1998)]{Kalb98} Kalberla, P.~M.~W., 
Westphalen, G., Mebold, U., Hartmann, D., \& Burton, W.~B.\ 1998, \aap, 
332, L61 


\bibitem[Kalberla(2003)]{Kalberla2003} Kalberla, P.~M.~W.\ 2003, 
\apj, 588, 805 


\bibitem[Kalberla et al.(2005)]{Kalberla2005} Kalberla, P.~M.~W., 
Burton, W.~B., Hartmann, D., Arnal, E.~M., Bajaja, E., Morras, R.,  
P{\"o}ppel, W.~G.~L.\ 2005, \aap, 440, 775 

\bibitem[Kalberla \& Haud(2006)]{Kalberla2006} Kalberla, P.~M.~W., 
\& Haud, U.\ 2006, \aap, 455, 481 

\bibitem[Kalberla et al.(2007)]{Kalberla2007} Kalberla, P.~M.~W., 
Dedes, L., Kerp, J., \& Haud, U.\ 2007, \aap, 469, 511 


\bibitem[Kuijken \& Gilmore(1989)]{KG89} Kuijken, K., Gilmore, G. 1989b, MNRAS 239, 651 


\bibitem[Kulkarni et al.(1982)]{Kulkarni1982} Kulkarni, S.~R., 
Heiles, C., \& Blitz, L.\ 1982, \apjl, 259, L63 

\bibitem[Kulkarni \& Fich(1985)]{Kulkarni1985} Kulkarni, S.~R., \& 
Fich, M.\ 1985, \apj, 289, 792 


\bibitem[Levine et al.(2006a)]{Levine2006a} Levine, E.~S., Blitz, 
L., \& Heiles, C.\ 2006a, \apj, 643, 881 

\bibitem[Levine et al.(2006b)]{Levine2006b} Levine, E.~S., Blitz, 
L., \& Heiles, C.\ 2006b, Science, 312, 1773 


\bibitem[Lockman(1988)]{Lockman1988} Lockman, F.~J.\ 1988, The 
Outer Galaxy, 306, 79 

\bibitem[Lockman \& Gehman(1991)]{Lockman91} Lockman, F.\ J.\ \& Gehman, C.\   S.\ 1991, \apj, 382, 182

\bibitem[Lockman(2002a)]{Lockman2002a} Lockman, F.~J.\ 2002, Seeing 
Through the Dust: The Detection of HI and the Exploration of the ISM in 
Galaxies, 276, 107 


\bibitem[Lockman(2002b)]{Lockman2002b} Lockman, F.~J.\ 2002, \apjl, 
580, L47 


\bibitem[Lockman(2004)]{Lockman2004} Lockman, F.~J.\ 2004, 
Recycling Intergalactic and Interstellar Matter, 217, 130 


\bibitem[McClure-Griffiths et al.(2006)]{McClure-Griffiths2006} 
McClure-Griffiths, N.~M., Ford, A., Pisano, D.~J., Gibson, B.~K., 
Staveley-Smith, L., Calabretta, M.~R., Dedes, L., 
\& Kalberla, P.~M.~W.\ 2006, \apj, 638, 196 


\bibitem[Maloney(1993)]{Maloney1993} Maloney, P.\ 1993, \apj, 414, 
41 


\bibitem[Mohan et al.(2004)]{Mohan2004} Mohan, R., Dwarakanath, 
K.~S., \& Srinivasan, G.\ 2004, Journal of Astrophysics and Astronomy, 25, 
185

\bibitem[M{\"u}nch \& Zirin(1961)]{Muench1961} M{\"u}nch, G., \& 
Zirin, H.\ 1961, \apj, 133, 11 
 

\bibitem[Nakanishi \& Sofue(2003)]{Nakanishi2003} Nakanishi, H., \& 
Sofue, Y.\ 2003, \pasj, 55, 191 

\bibitem[Oosterloo et al.(2007)]{Oosterloo2007} Oosterloo, T., 
Fraternali, F., \& Sancisi, R.\ 2007, \aj, 134, 1019 

Lockman, F.~J., \& Shields, J.~C.\ 2007, \apj, 656, 928 

\bibitem[Pikelner \& Shklovsky(1958)] {Pikelner1958} Pikelner S.B., Shklovsky
  I.S., 1958, IAU Symposium 8, Reviews of Modern Physics, 30, 935.


\bibitem[Radhakrishnan \& Srinivasan(1980)]{Radhakrishnan1980} 
Radhakrishnan, V., \& Srinivasan, G.\ 1980, Journal of Astrophysics and 
Astronomy, 1, 47 


\bibitem[Rohlfs(1977)]{Rohlfs1977} Rohlfs, K.\ 1977, Lecture Notes 
in Physics, Berlin Springer Verlag, 69,  

\bibitem[Saha \& Jog(2006)]{Saha2006} Saha, K., \& Jog, C.~J.\ 
2006, \mnras, 367, 1297 

\bibitem[Spitzer(1942)]{Spitzer42}Spitzer, L.,  1942, ApJ 95, 329

\bibitem[Spitzer(1956)]{Spitzer1956} Spitzer, L.~J.\ 1956, \apj, 
124, 20 

\bibitem[Stanimirovi{\'c} et al.(2006)]{Stanimirovic2006} 
Stanimirovi{\'c}, S., et al.\ 2006, \apj, 653, 1210 

\bibitem[Stark et al.(1992)]{Stark1992} Stark, A.~A., Gammie, 
C.~F., Wilson, R.~W., Bally, J., Linke, R.~A., Heiles, C., \& Hurwitz, M.\ 
1992, \apjs, 79, 77 


\bibitem[Stil et al.(2006)]{Stil2006} Stil, J.~M., et al.\ 2006, \apj, 637,
  366

\bibitem[Swaters et al.(1997)]{Swaters1997} Swaters, R.~A., 
Sancisi, R., \& van der Hulst, J.~M.\ 1997, \apj, 491, 140 

\bibitem[Twarog(1980)]{Twarog1980} Twarog, B.~A.\ 1980, \apj, 242, 
242 


\bibitem[van Woerden et al.(2004)]{HVCbook} van Woerden, H., 
Wakker, B.~P., Schwarz, U.~J., \& de Boer, K.~S.\ 2004a, ASSL Vol.~312: High 
Velocity Clouds


\bibitem[Voskes(1999)]{Voskes1999} Voskes, T. 1999, M.Sc. Thesis, University of Leiden, 
astro-ph/0601653 

\bibitem[Wakker(1991)]{Wakker1991} Wakker, B.~P.\ 1991, \aap, 250, 499 


\bibitem[Weinberg(1998)] {Weinberg1998} Weinberg, M.~D.\ 1998, 
\mnras, 299, 499 

\bibitem[Weiner \& Sellwood(1999)]{Weiner99} Weiner, B.~J.~\& Sellwood, J.~A.\ 1999, \apj, 524, 112

\bibitem[Wolfire et al.(2003)]{Wolfire2003} Wolfire, M.~G., McKee, C.~F.,
  Hollenbach, D., Tielens, A.~G.~G.~M., 2003, \apj, 587, 278

\bibitem[Wouterloot et al.(1990)]{Wouterloot90} 
Wouterloot, J.~G.~A., Brand, J., Burton, W.~B., \& Kwee, K.~K.\ 1990, \aap, 
230, 21 

\end{thebibliography}
\end{document}